\documentclass[aps,twocolumn,prb,floatfix,showpacs,superscriptaddress]{revtex4-2}
\usepackage{amsmath}
\usepackage{amssymb}
\usepackage{amsfonts}
\usepackage[pdftex]{graphicx}
\usepackage{units}
\usepackage[usenames]{color}
\usepackage{dcolumn}
\usepackage{bm}
\usepackage{float}
\usepackage[utf8]{inputenc}
\usepackage[colorlinks=true,citecolor=blue,linkcolor=blue]{hyperref}
\usepackage[normalem]{ulem}
\usepackage{xcolor}
\usepackage{makecell}
\usepackage{changes} 

\newcommand{\br}{\bm{r}}

\newcommand{\bE}{\bm{E}}

\renewcommand{\mathbf}[1]{\ensuremath{\boldsymbol{#1}}}

\begin{document}

\title{Real-time simulations of laser-induced electron excitations in crystalline ZnO}

\author{Xiao Chen}
\affiliation{Research Center Future Energy Materials and Systems of the Research Alliance Ruhr, Germany} 
\affiliation{Faculty of Physics and Astronomy and ICAMS, Ruhr University Bochum, Universitätstrasse 150, 44780 Bochum, Germany}
\affiliation{Institut für Festkörpertheorie und -optik, Friedrich-Schiller-Universität Jena, Max-Wien-Platz 1, 07743 Jena, Germany}
\affiliation{European Theoretical Spectroscopy Facility}
\author{Thomas Lettau}   
\affiliation{Institut für Festkörpertheorie und -optik, Friedrich-Schiller-Universität Jena, Max-Wien-Platz 1, 07743 Jena, Germany}
\author{Ulf Peschel}
\affiliation{Institut für Festkörpertheorie und -optik, Friedrich-Schiller-Universität Jena, Max-Wien-Platz 1, 07743 Jena, Germany}
\author{Nicolas Tancogne-Dejean}  
\affiliation{Max Planck Institute for Structure and Dynamics of Matter and Center for Free-Electron Laser Science, Hamburg, 22761, Germany}
\affiliation{European Theoretical Spectroscopy Facility}
\author{Silvana Botti}  
\affiliation{Research Center Future Energy Materials and Systems of the Research Alliance Ruhr, Germany} 
\affiliation{Faculty of Physics and Astronomy and ICAMS, Ruhr University Bochum, Universitätstrasse 150, 44780 Bochum, Germany}
\affiliation{Institut für Festkörpertheorie und -optik, Friedrich-Schiller-Universität Jena, Max-Wien-Platz 1, 07743 Jena, Germany}
\affiliation{European Theoretical Spectroscopy Facility}
\date{\today}

\begin{abstract}  
We investigate non-equilibrium electron dynamics in crystalline ZnO induced by ultrashort, relatively intense, infrared laser pulses. Our focus is on understanding the mechanism that facilitates efficient conduction band population in ZnO to enable optically pumped lasing. We consider two different pulse frequencies (in the near-infrared and mid-infrared) for which experimental data are available, and we calculate the electronic response of a ZnO crystal for a wide range of pulse intensities. We apply and compare three complementary theoretical approaches: the analytical Keldysh model, the numerical solution of the semiconductor Bloch equations, and real-time time-dependent density functional theory. We conclude that time-dependent density functional theory is a valid \textit{ab initio} approach for predicting conduction band population, that offers an accurate enough description of static and transient optical properties of solids and provides physics insight into the intermediate excitation regime, where electronic excitations
are determined by the interplay of intraband tunneling, a consequence of band bending, and interband multi-photon absorption.
\end{abstract}

\maketitle   

\section{Introduction}  

ZnO nanolasers are promising systems for applications in optoelectronics, offering electrically pumped laser capabilities that can be utilized in various fields such as optical communication, biosensing, medical imaging, and 3D displays\cite{ZnOnanowire1,ZnOnanowire2,ZnOnanowire3,ZnOnanowire4}.

Since the initial report of lasing in ZnO in 1966~\cite{1966firstZnOlaser}, over the subsequent decades, ZnO has been grown in various forms, including macroscopic single crystals~\cite{znomacro}, thin films~\cite{ZnOfilm}, quantum wells~\cite{ZnOwell}, quantum rods (nanowires)~\cite{2001firstZnOnanowire}, and quantum dots~\cite{ZnOdot}, all exhibiting substantial luminescence. Nanowires have attracted increased attention, as they fulfill all the requirements for realizing a nanolaser: ZnO serves as a well-studied gain medium, being a direct bandgap material with a large bandgap energy of 3.3 eV at 300 K~\cite{zno3.3ev}, the end facets and cylindrical geometry of the nanowire precisely form a standalone Fabry-Pérot type resonance cavity, and the available optical pump mechanism proves particularly useful, especially for biological applications~\cite{bioapplication}. In 2001, room-temperature laser emission from ZnO nanowires was reported for the first time~\cite{2001firstZnOnanowire}.

While the potential of emerging ZnO nanolasers is promising, the key challenge lies in effectively injecting charge carriers to achieve sufficient gain and overcoming the significant optical losses inherent in such compact laser devices. A major drawback exists in the fundamental pumping mechanism: stimulated emission via linear optical processes in ZnO nanowires inherently requires intense, coherent deep-UV excitation sources. The excitation photon energy must surpass the band gap ($\sim$3.3 eV) of ZnO nanowires for one-photon absorption pumping. Relying on expensive deep-UV pumping sources poses significant challenges to the widespread application of ZnO nanolasers due to overall cost, complexity, and incompatibility with chip-level integration.

An alternative approach under investigation for ZnO nanowires relies on an emission process induced by multi-photon absorption (MPA)~\cite{mpa0,mpa1, mpa2}. When exposed to intense near-infrared (IR) radiation, the nonlinear interaction between the applied optical field and ZnO nanostructures results in the simultaneous absorption of two or more photons of sub-bandgap energy through a virtual-state assisted interband transition. This process generates electron-hole pairs, leading to band-edge emission via radiative recombination. Near-IR intense light pulses, readily produced with inexpensive laser diodes, thus offer a solution for chip-level design and low-cost implementation of ZnO nanolasers in photonic circuitries and sensing systems.

Recently, Hollinger {\it et al.} reported a new experimental realization of a ZnO nanowire laser~\cite{Hollinger2}, demonstrating that varying the wavelength of the pumping pulse does not decrease the lasing threshold intensity $I_{\rm th}$. In fact, by tuning the wavelength from the near-infrared at 0.8 $\mu m$ (1.55 eV) to the mid-infrared at 3.5 $\mu m$ (0.35 eV), the threshold intensity $I_{\rm th}$ changed only from 5$\times10^{15}$ to 6$\times10^{15}$ W/m$^2$. This observation suggests that the underlying electronic excitation mechanism undergoes modifications with variations of both pumping wavelength and intensity.

There are different regimes to describe electronic excitations in semiconductors. The first is the already mentioned single-photon absorption or linear-response regime, which requires a pumping photon energy $\hbar\omega_p$ larger than the bandgap energy $E_{\rm g}$. It can be achieved with low intensities, long pump durations, and even continuous-wave pulses. Another regime is possible when $\hbar\omega_p < E_{\rm g}$ and the pump intensity is sufficiently high, involving non-linear electronic excitations. The dominant processes can be either multi-photon absorption (MPA)~\cite{mpa0,mpa1, mpa2} or electron tunneling~\cite{tunneling, tunneling1}. 
The latter is dominant for small values of $\hbar\omega_p$ and high pump intensity. In this case, the strong laser field bends the electronic bands, enabling direct tunneling of electrons from the valence bands to the conduction bands. Thus, the transition rate becomes independent of the photon energy, but solely depends on the intensity of the pumping light. In the lower intensity limit, MPA is a perturbative nonlinear process, where the strength of absorption of N-photons scales as the N-th power of the laser intensity. Particularly interesting is the intermediate regime in which both the multiphoton-driven and field-driven mechanisms significantly contribute to the population of the conduction band.

Understanding the evolution of the electronic excitation mechanism due to varying pumping wavelengths and intensities holds profound implications for the optimization and practical utilization of ZnO nanowire lasers. Our study offers comprehensive insight into this phenomenon through the application of multiple complementary computational methods to elucidate the various excitation mechanisms. By comparing our calculations with experimental results, we can explain the observed variations in threshold intensities across different pulse wavelengths. This comparison enhances the fundamental understanding of ZnO nanowire laser operation.

The structure of this paper is as follows. Section II provides an introduction to the theoretical approaches: the Keldysh model (A), the semiconductor Bloch equations (SBE) (B), and time-dependent density functional theory (TDDFT) (C). Section IIIA presents simulations of electron dynamics evolution under excitation pulses of varying intensities. Comparative analyses of simulations for near-infrared and mid-infrared pulses are detailed in Sections IIIB and IIIC, respectively, while Section IIID discusses the dynamical Franz-Keldysh effect for mid-infrared pumping. Further details on the calculation of dipole matrix elements, the impact of dephasing time on the SBE solution, the influence of bandgap errors on the SBE and Keldysh model, and the effect of shortened propagation time in TDDFT are elaborated in the supplemental material (SM). Throughout the paper, SI units are employed.

\section{Theory background and methods}

\subsection{Keldysh model}

The Keldysh model, introduced by L.V. Keldysh in 1964 ~\cite{keldysh}, is a widely employed analytical theory based on non-equilibrium Green's functions to calculate light-induced excitation rates across a broad wavelength range. Keldysh originally developed this model to evaluate non-relativistic ionization rates of hydrogen-like atoms and two-band solids subjected to strong laser fields, described in the length gauge and within the dipole approximation. 
To apply this approach to solids we rely on a two-band model in which the modifications of the band structure induced by the laser is translated into an effective bandgap that takes into account the Stark-shift of the energy levels and can therefore describe below-bandgap photon-assisted excitation~\cite{Keldysh2014}. 

According to the Keldysh model, MPA is the dominant process, for excitations at photon energies smaller than the band gap, when the dimensionless Keldysh parameter $\gamma = \omega \sqrt{m^{\star} E_g} / {e E}$ ~\cite{Siiman2009} is large ($\gamma \gg 1$), while electron tunneling is the dominating mechanism for  $\gamma \ll 1$. In the formula for $\gamma$, $\omega$ is the pump laser frequency, $m^{\star}$ is the effective mass in units of the electron mass $m_e$ ($m^{\star} = 1.88 m_e$ for the lowest conduction band of ZnO at the $\Gamma$ point~\cite{ZnO_mass}), $e$ is the absolute value of the electron charge and $E$ is the laser electric field strength.  

The strength of the electric field E is related to the  pump laser intensity I, 
\begin{equation}
I = \frac{1}{2} n \epsilon_0 c E^2 \,, 
\label{intensity}
\end{equation}
where $\epsilon_0$ is the vacuum dielectric permittivity, $c$ is the vacuum speed of light, and the refractive index $n$ is equal to 1  in air. 

The excitation rate $W^{\rm Keldysh}_{\rm ex}$ in units of $m^{-3} \cdot s^{-1}$ denotes the number of excited electrons per unit volume and time, and it can be calculated using the following equations~\cite{keldysh, Siiman2009},
\begin{equation}
\begin{aligned}
    W^{\rm Keldysh}_{\rm ex} &= 
    \frac{2 \omega}{9 \pi} 
    \left( 
    \frac{\sqrt{1+\gamma^2}}{\gamma} \frac{m^{\star} \omega}{\hbar} \right)^{3/2}
    Q(\gamma, \frac{\tilde{E}_g}{\hbar \omega}) \\
    &\times {\exp}{\left(\frac{- \pi k [K(\gamma_1)-E(\gamma_1)]}{E(\gamma_2)}\right)}\,,
\end{aligned}
\end{equation}
with the dimensionless Coulomb factor
\begin{equation}
\begin{aligned}
Q(\gamma,x) &=  \left(\frac{\pi}{2K(\gamma_2)}\right)^{1/2}  \\
&\times \sum_{n=0}^{\infty} \{ {\exp}(\frac{-\pi n[K(\gamma_1)-E(\gamma_1)]}{E(\gamma_2)})  \\
&\times \Phi( [\frac{ \pi^2(2\lfloor x+1 \rfloor-2x +n ) }{ 2 K(\gamma_2) E(\gamma_2) }]^{1/2} ) \} \,,
\end{aligned}
\end{equation}
the renormalized bandgap that considers the energy shift induced by the Stark effect
\begin{equation}
\tilde{E}_g = \frac{2}{\pi} E_g \frac{\sqrt{1+\gamma^2}}{\gamma} E(\gamma_2) \,,
\label{Eg}
\end{equation}
and the Dawson integral
\begin{equation}
\Phi(x) = \int^x_0 {\exp}(y^2 - x^2) dy \,.
\end{equation}

In the above equations, the parameters $\gamma_1 = \gamma / \sqrt{1 + \gamma^2}$ and $\gamma_2 = 1 / \sqrt{1 + \gamma^2}$ are dimensionless, like $\gamma$. $K(x)$ and $E(x)$ are complete elliptic integrals of the first and second kind, respectively:

\begin{equation}
\begin{aligned}
& K(x) =  \int^{\pi/2}_0  (1- x^2  {\sin}^2\theta) ^{-1/2}  d\theta \,, \\
& E(x) =  \int^{\pi/2}_0  (1- x^2  {\sin}^2\theta) ^{1/2}  d\theta \,.
\end{aligned}
\end{equation}
We can define the order of the MPA using the parameter $k$:
\begin{equation}
k = \lfloor \frac{\tilde{E}_g}{\hbar \omega} + 1 \rfloor \,,  
\label{eq:k}
\end{equation}

where $\lfloor... \rfloor$ denotes rounding down to the nearest integer. Once the excitation rate $W^{\rm Keldysh}_{\rm ex}$ is calculated, we can determine the population of the conduction band induced by a laser pulse with a specific intensity $I_{\rm p}$ and central frequency $\omega$, when the medium is a crystalline material with a band gap $E_{\rm g}$.

We note that the Keldysh model has some limitations: it is valid for not too strong laser fields and photon energies smaller than the band gap, it does not take actual band structures into account~\cite{Keldysh2014,Keldysh1}, and it overlooks electron-electron interactions and many-body effects, so we don't expect it to work well in all situations. 

\subsection{Semiconductor Bloch Equations}

To address the limitations of the Keldysh model, we turn to the semiconductor Bloch equations (SBE), which offer significant advantages in capturing complex carrier dynamics, multiband effects, and many-body effects. SBE can take into account the calculated band structure of ZnO, allowing for a precise description of transitions and coupling effects, while  the Keldysh model provides a simpler analytical description. Furthermore, dephasing effects can be included in the framework of the SBE~\cite{Lindberg1988}.

One way~\cite{Yue2022} to derive the SBE  is by starting with the Hamiltonian ${H}_{\text{SBE}}$ that describes electrons in a crystalline potential with a light field $\bE(t)$ coupled to the electronic system in the dipole approximation.

\begin{equation}
{H}_{\text{SBE}}(\br,t) = {T} + V(\br) + \br \cdot e\bE(t) \,,
\label{Hsbe}
\end{equation}

The wavefunction $\Psi(\br, t)$ is the solution of the time-dependent Schrödinger equation (TDSE), 

\begin{equation}
i \hbar \frac{\partial\Psi(\br,t)}{\partial t } = {H}_{\text{SBE}}(\br,t) \Psi(\br,t),
\label{TDSE}
\end{equation}

can be expanded into the basis of the time-independent Bloch states $\phi_n^{\mathbf{k}}(\br)$ with time-dependent coefficient $a_n^{\mathbf{k}}(t)$~\cite{Yue2022}, where the indices indicate the band index  and $\mathbf{k}$-point,

\begin{equation}
\Psi(\br, t) =\sum_{n,\mathbf{k}\in \text{BZ}}a_n^{\mathbf{k}}(t)\phi_n^{\mathbf{k}}(\br). 
\label{psi}
\end{equation}

Then, we can rewrite the TDSE Eq.~\eqref{TDSE} using Eq.~\eqref{psi} and  project onto the Bloch states,  obtaining the equation of motion for the time-dependent density matrix 
$\hat{\rho}(t) = |\Psi(\br, t)\rangle \langle\Psi(\br, t)| $ 
with matrix elements 
$\rho_{mn}^{\mathbf{k}}(t) = a_m^{\mathbf{k}}(t){a_n^{\mathbf{k}}}^{*}(t)$ ~\cite{Yue2022}:

\begin{equation}
\begin{aligned}
       & i \hbar \frac{d{\rho}_{mn}^{\mathbf{k}}(t)}{dt} =  (E_m^{\mathbf{k}} - E_n^{\mathbf{k}}) \rho_{mn}^{\mathbf{k}}(t) 
          - i \hbar (1 -\delta_{mn}) \rho^{\mathbf{k}}_{mn}(t)/T_2\\ 
        &      + i  e\mathbf{E}(t) \cdot \nabla_{\mathbf{k}}\rho^{\mathbf{k}}_{mn}(t)
        +  e\mathbf{E}(t) \cdot \sum_{l}
        \left[ \mathbf{d}^{\mathbf{k}}_{ml}\rho^{\mathbf{k}}_{ln}(t) - \mathbf{d}^{\mathbf{k}}_{ln}\rho^{\mathbf{k}}_{ml}(t)   
        \right] ,
\label{EOM}
\end{aligned}
\end{equation}

where $E_m^{\mathbf{k}}$ is the band energy and $\mathbf{d}_{mn}^{\mathbf{k}}$ is the dipole matrix element for transitions between the bands $m$ and $n$ with $n\neq m$, or the Berry connections for $m=n$. 
The band structure and dipole matrix elements can be obtained from knowledge of the Kohn-Sham energies and wavefunctions, calculated for ZnO within DFT+U, as detailed in Sec.II.C. 

We want to simulate the experimental pump-laser pulses used in Ref.~\cite{Hollinger2}: a 30 fs pulse with a threshold intensity $I_{\rm th}^{0.8} = 5 \times 10^{15}$ W/m$^2$ and a near-IR central wavelength of 0.8 $\mu$m and a 105 fs pulse with $I_{\rm th}^{3.5} = 6\times 10^{15}$ W/m$^2$ and a mid-IR central wavelenght of 3.5 $\mu$m. We use the same pulse shape, but reduce the pulse duration to make TDDFT calculations feasible. The validity of this approximation is discussed in Section S4 of the SM, where we show that reducing the propagation time does not lead to significant changes in the final values. 

The shape of the simulated laser field outside the ZnO crystal is described by Eq.~\eqref{laser},

\begin{equation}
\mathbf{E}(t) =
\begin{cases}
 E_0 \sin(\omega t ) \sin^2(\pi t / T) \hat{e_z}     &\rm{if} \  0<t<T \\
 0   &\rm{else}.
\end{cases}
\label{laser}
\end{equation}

Here $E_0$ is the strength of the electric field, $\omega$ is the laser frequency, $T$ is the pulse duration, the polarization direction $\hat{e}_z$ is aligned with the $c$ axis of the hexagonal ZnO structure.
For the  near-IR 0.8 $\mu$m laser, we use a simulated pulse duration  of  T = 12.1 fs, and the total propagation time is 14 fs. 
For the mid-IR 3.5 $\mu$m laser, the pulse duration  is T = 31.5 fs, and the total propagation time is 36.3 fs. 
The total propagation time must be longer than the pulse duration to read out a stable excitation electron number after the pulse ends.
The same laser field is used for SBE and TDDFT calculations.  

To describe the many-body effects, like electron-electron, electron-hole and electron-phonon scattering, we introduce a phenomenological dephasing term  at the right side of Eq.~\eqref{EOM} with dephasing time $T_2$~\cite{T2}, which describes all the incoherent loss processes involved.  In literature, a commonly employed approach consist in defining $T_2$ as a  fraction of an optical cycle ($T_2 \approx T_0 / 4$~\cite{T2:1}) or about 1 fs~\cite{T2:2,T2:3}. Such short $T_2$  raises important  conceptual  questions concerning the ultrafast decoherence processes for electronic excitations in solids~\cite{T2}.  

The choice of $T_2$ significantly affects the excited electron density $n_{\rm ex}$, which varies from $2\times 10^{25}/m^3$ to $6\times10^{24}/m^3$ as $T_2$ increases from 10 fs to infinity. The effect of $T2$ on the excitation densities are discussed in Section S2 of the SM. Here we decide to set $T_2$ = 20 fs as we find it is a good compromise.

We note that there exist more complicated formulations of the SBE, including explicitly Coulomb interaction and excitonic effects~\cite{KIRA2006155}. In this work, these effects are not included. 

\begin{figure}[htbp!]
\centering
\includegraphics[width = 0.9\columnwidth]{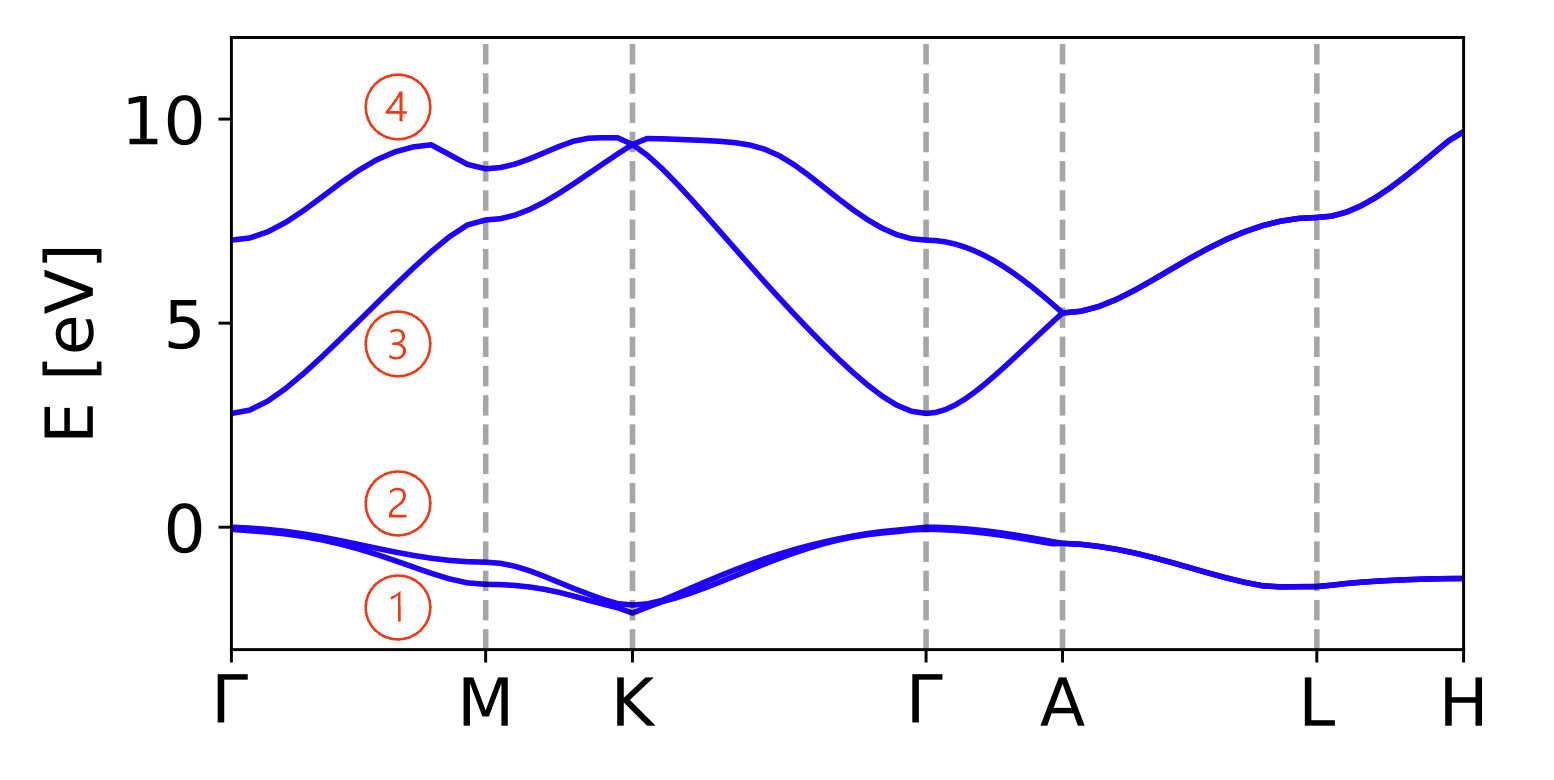}
\caption{Band structure of bulk ZnO computed for SBE using a 4-bands model. We label the two valence bands as 1 and 2, and the two conduction bands as 3 and 4.} 
\label{band} 
\end{figure}

We approximate the band structure of ZnO using a four-bands model (4BM). Fig.~\ref{band} illustrates the band structure $E^{\mathbf{k}}_{m}$ of ZnO. In this figure, we show two valence bands (VBs), labeled as 1 and 2, and two conduction bands (CBs), labeled as 3 and 4, that constitute our 4BM. The calculations of $E^{\mathbf{k}}_m$ and $\mathbf{d}_{mn}^{\mathbf{k}}$ were conducted using the DFT+U method with a $12 \times 12 \times 41$ k-points sampling grid across the entire Brillouin zone.
The presence of degenerate bands along high-symmetry directions of the Brillouin zone for both VBs and CBs can lead to inaccuracies when using a simplified two-bands model (2BM), as the dipole matrix elements of degenerate levels can be very different (see Fig.1S of the SM). Employing a 4BM significantly mitigates these errors.

Given that the external laser is $z$-polarized, we only need to consider the $z$-components of the dipole matrix elements $\mathbf{d}_{mn}^{\mathbf{k}}$ (calculated matrix elements are shown in Section S1 of the SM). 

When calculating the gradient term $\nabla_{\mathbf{k}}\rho^{\mathbf{k}}_{mn}(t)$ in Eq.~\eqref{EOM}, we enhance the resolution along the $k_z$ direction by employing Akima 1D interpolation~\cite{Interpolation} with periodic boundary conditions for both $E(\mathbf{k})$ and $\mathbf{d}_{mn}^{\mathbf{k}}$. This method utilizes a continuously differentiable sub-spline constructed from piecewise cubic polynomials. The interpolation increases the number of $k_z$ points from 41 to 300, corresponding to a spacing of $\Delta k_z = 0.04 \ \text{nm}^{-1}$.
 
For the initial condition of propagation, we set $\rho^{\mathbf{k}}_{11}(t=0) = \rho^{\mathbf{k}}_{22}(t=0) = 1$, with all other elements being zero.
To solve the differential equations Eq.~\eqref{EOM}, we choose the 12th-order Adams-Bashforth method~\cite{Adams–Bashforth} due to its high accuracy and efficiency, the time step $\Delta t$ is set as $10^{-3}$ fs.

Then, we can calculate the density matrix elements for each time step. The number of electrons excited from VBs to CBs, $n^{\rm SBE}_{\rm ex}(t)$, is obtained as follows:

\begin{equation}            
    n^{\rm SBE}_{\rm ex}(t) 
     =  \frac{2}{V} \sum_{\mathbf{k}} w_{\mathbf{k}} (\rho_{33}^{\mathbf{k}}(t)  +\rho_{44}^{\mathbf{k}}(t) ) \,,
\label{nsbe}
\end{equation}

where $V$ is the unit cell volume, $w_{\mathbf{k}}$ is the $\mathbf{k}$-point weight, the summation of $\mathbf{k}$-points is over the whole irreducible  Brillouin zone, and the factor of 2 accounts for the spin of the electrons.

\subsection{Time-dependent density functional theory}

Time-dependent density functional theory (TDDFT) is an extension of DFT designed to calculate the properties of many-electron systems under the influence of time-dependent external potentials, such as electric or magnetic fields. Real-time TDDFT stands out as one of the most appealing methods for investigating electron dynamics under ultrashort laser pulse irradiation,  enabling studies beyond linear response.  The formal foundation of TDDFT is the Runge-Gross theorem ~\cite{Runge} that establishes a unique mapping between  the time-dependent external potential of a system and its time-dependent density $n(\boldsymbol{r},t)$  for a given initial wavefunction.  The $n(\boldsymbol{r},t) = \sum_{n,\mathbf{k}\in \text{BZ}} | \psi_{n,\mathbf{k}} (\boldsymbol{r},t)|^2$ is determined by solving an auxiliary set of Schrödinger equations, the time-dependent Kohn-Sham (TDKS) equations,

\begin{equation}   i \hbar \frac{\partial \psi_{n,\mathbf{k}}(\boldsymbol{r},t)}{\partial t} = H_{\rm KS}(\boldsymbol{r},t)\psi_{n,\mathbf{k}}(\boldsymbol{r},t).
\label{eq_tdks}
\end{equation}	

Here, the indices $n$ and $\mathbf{k}$ indicate the band index and $\mathbf{k}$-point, respectively. $\psi_{n,\mathbf{k}}(\boldsymbol{r},t)$ is the time-dependent single-particle wave function. $H_{\rm KS}(\boldsymbol{r},t)$ is the time-dependent Kohn-Sham Hamiltonian, and it can be written in velocity gauge~\cite{oscillation}, as shown in Eq.~\eqref{ks}. The gauge transformation is employed to circumvent discontinuities arising from the electric field propagation in periodic systems~\cite{gauge}, 

\begin{equation}
\begin{aligned}
     H_{\rm KS}(\boldsymbol{r},t) =  &\frac{1}{2m_e} (\boldsymbol{p} + \frac{e}{c} \boldsymbol{A}_{\rm tot}(t))^2 + V_{\rm ion}(\boldsymbol{r},t)  \\
     & + V_{\rm H}(\boldsymbol{r},t) + V_{\rm xc}(\boldsymbol{r}, t).
\end{aligned}
\label{ks}
\end{equation}

Here, the first term on the right side $\frac{1}{2m_e} (\boldsymbol{p} + \frac{e}{c} \boldsymbol{A}_{\rm tot}(t))^2$ includes the kinetic energy of electrons and the laser field, described in the velocity gauge,  $m_e,e,c,\boldsymbol{p}$ denote the electron mass, elementary charge, vacuum light speed, and electron momentum, respectively. The potential terms $V_{\rm ion}$, $V_{\rm H}$ and $V_{\rm xc}$ indicate the pseudopotential, the Hartree potential and the exchange-correlation potential, respectively. We omitted here the nonlocal contribution to the external potential from the pseudopotentials for simplicity. 

The total vector potential $\boldsymbol{A}_{\rm tot}(t)$ is time-dependent and spatially uniform. It is composed of the external and induced vector potentials: 
$\boldsymbol{A}_{\rm tot}(t) = \boldsymbol{A}_{\rm ext}(t) + \boldsymbol{A}_{\rm ind}(t)$. The external vector potential $\boldsymbol{A}_{\rm ext}(t)$ is related to the electric field of the applied laser pulse $\boldsymbol{E}_{\rm ext}(t)$ by 

\begin{equation}
\boldsymbol{E}_{\rm ext}(t) = - \frac{1}{c} \frac{d \boldsymbol{A}_{\rm ext}(t)}{dt}\,.
\label{EA}
\end{equation}

where $\boldsymbol{E}_{\rm ext}(t)$ is described by Eq.~\eqref{laser}.
The same notation is applied to the total electric field $\boldsymbol{E}_{\rm tot} = \boldsymbol{E}_{\rm ext} + \boldsymbol{E}_{\rm ind}$.
The induced macroscopic vector potential is given by the Maxwell equation  

\begin{equation}
 \frac{\partial^2}{\partial t^2}\mathbf{A}_{\mathrm{ind}}(t) = \frac{4\pi c}{V}\mathbf{j}(t)\,,
 \label{eq:maxwell}
\end{equation}
with cell volume V and macroscopic current $\mathbf{j}(t)$~\cite{NiO,j2A1}.

As we perform TDDFT simulations of electron dynamics in a bulk ZnO crystal, we cannot treat explicitly surface-induced electric fields and we need therefore to redefine consistently the intensity of the total field inside and outside ZnO. 
We follow the approach explained and validated in Refs.~\cite{Keldysh1,E_ind}. 
There two different ways to proceed that give approximately the same. We consider and compare both, calling them ``modified'' and ``unmodified'' calculations. A detailed description of these approximations can be found in Sec. III.A.


The second term of Eq.~\eqref{ks}, $V_{ion}(\boldsymbol{r},t)$, describes the electron-ion interaction. We employ Pseudodojo PBE  norm-conserving pseudopotentials~\cite{pseudodojo}
to describe the interaction between ions and valence electrons. These pseudopotentials accounts for the 6 valence electrons of the oxygen atom ($2s^2 2p^4$) and 20 valence electrons of the zinc atom ($3s^2 3p^6 3d^{10} 4s^2$). The energy transfer from electrons to ions is ignored by freezing the ionic position.

The third term $V_H(\boldsymbol{r},t) =  e^2 \int d \boldsymbol{r}^{\prime} \frac{n(\boldsymbol{r},t)}{|\boldsymbol{r} - \boldsymbol{r}^{\prime}|}$ is the Hartree potential, which describes the classical part of the electron-electron interaction.

The forth term $V_{xc}(\boldsymbol{r}, t)$ is the exchange-correlation potential.  In this work, we utilize the adiabatic generalized gradient approximation functional  proposed by Perdew, Burke and Ernzerhof (GGA-PBE) in 1997~\cite{pbegga1}.  It is acknowledged that the GGA Kohn-Sham band gap significantly underestimates the experimental quasiparticle gap. In this case, the calculated GGA-PBE direct band gap is 0.79 eV, while the experimental direct bandgap value is 3.3 eV~\cite{zno3.3ev}. 
The appropriate bandgap value is crucial for the excitation properties calculations. To open the gap, we employ  the DFT+U method, which is extended to the TDDFT scheme in Ref.~\cite{DFT+Uoctopus}, resulting in an improved Kohn-Sham bandgap energy of 2.81 eV. The  fundamental  idea behind DFT+U is to address the strong on-site Coulomb interaction of localized electrons by introducing an  additional Hubbard-like term. We include the Hubbard energy  $E_U$ in the DFT total energy, and $E_U$ is dependent solely on the   effective Hubbard U parameter $U_{\rm eff}$.  This parameter can be extracted from ab-initio calculations or set empirically. In this work, we adopt empirical values of  $U_{\rm eff}$ = 12.8 eV for $3d$ orbitals of zinc atom and $U_{\rm eff}$ = 5.29 eV for $2p$ orbitals of oxygen atom from Ref.~\cite{plusU}.  Although the $U_{\rm eff}$ applied on the $p$ electrons is not so well justified from a theoretical perspective, it is a common choice to obtain an improved bandgap.

The number of electrons excited from the valence to the conduction bands $n^{\rm TDDFT}_{\rm ex}(t)$ can be calculated in the TDDFT framework~\cite{oscillation} as

\begin{equation}
n^{\rm TDDFT}_{\rm ex}(t) = \frac{2}{V}  \sum_{nn\prime \mathbf{k}} w_{\mathbf{k}} ( \delta_{nn^{\prime}} - \left| \langle \phi_{n\mathbf{k}}|\psi_{n^{\prime}\mathbf{k}}(t) \rangle \right|^2 )\,.
\label{ntddft}
\end{equation}

Here $n$, $n'$ are the indices, $\mathbf{k}$ is the Bloch wave number, $w_{\mathbf{k}}$ is the $\mathbf{k}$-point weight, $\phi_{n\mathbf{k}}$ is the ground state Kohn-Sham wave function, $\psi_{n'\mathbf{k}}(t)$ is the time-evolved wave function, $\rm V = 49.668 \mathring{A}^3$ is the volume of unit cell, and the factor of 2 accounts for the spin of the electrons.

The DFT and TDDFT calculations were performed in the Kohn-Sham scheme using $\textbf{OCTOPUS}$~\cite{octopus, octopus1}, a real-time, real-space package that makes use of pseudopotentials to simulate the electron-ion dynamics of one, two, and three-dimensional finite and periodic systems subject to time-dependent electromagnetic fields.

Crystalline ZnO possesses a hexagonal wurtzite crystal structure, falling within the  $P6_3mc$ space group (\# 186). The primitive cell comprises 2 zinc atoms and 2 oxygen atoms.  The optimized  lattice parameters, calculated within DFT with the GGA-PBE functional~\cite{pbegga1}, are $a=b=3.289\mathring{A}$, $c = 5.307\mathring{A}$, $\alpha=\beta=90^{\circ}$, and $\gamma = 120^{\circ}$, exhibiting a $1.21\%$ deviation from the room-temperature x-ray-diffraction experimental data~\cite{ZnOexplattice}: $a=b=3.250\mathring{A}$, $c=5.204\mathring{A}$. This results in a DFT primitive cell volume of $49.668 \mathring{A}^3$. Periodic boundary conditions were applied to the primitive cell. 

The wave functions, densities, and potentials were discretized with a real-space mesh grid with spacing 0.1 $\mathring{A}$ which is equivalent to the grid cutoff energy $E_{\rm cut}$ = 3.3 keV. The $\mathbf{k}$-point sampling method in reciprocal space  followed the Monkhorst-Pack scheme~\cite{ksampling}, the grid was set as $12 \times 12 \times 7$, with convergence precision of 4 meV/atom, which corresponds to 1008 $\mathbf{k}$-points in the whole Brillouin zone. 
 
The external laser field in Eq.~\eqref{laser} was produced to impinge on the simulation box with the polarization along the $z$-direction,  breaking the symmetry in this direction. So we can utilize the remaining 6 symmetries of the system, and reduce the $\mathbf{k}$-points sampling number from 1008 to 217.  The time step is chosen as  $\Delta t  =  3.6 \times 10^{-4}$ fs after thorough convergence tests. 

The propagator algorithm employed the enforced time-reversal symmetry method~\cite{propagator},  where the wave function at time step $(n+1)$ can be calculated by the information of  time step $n$: $\psi_{n+1} = \exp(-i\delta t H_{n+1} / 2) \exp(-i \delta t H_n / 2 ) \psi_n$. The Hamiltonian $H_{n+1}$ is estimated by a single exponent $\psi_{n+1}^{\star} = \exp(-i \delta t H_n) \psi_n$ and $H_{n+1} = H[\psi_{n+1}^{\star}]$. We choose the default 4th order Taylor expansion for the time-evolution operator, $\exp(-i \delta t H) = \sum_{i=0}^4 \frac{(-i \delta t)^i}{i!} H^i$, as recommended in Ref.~\cite{4thtaylor}. Since the Kohn-Sham propagator is nonlinear, for each propagation step we perform the self-consistent correction to decrease the propagation error, using a self consistency accuracy threshold of 0.03 meV.

We note that the SBE can be connected with TDDFT. Turkowski and coworkers~\cite{PhysRevB.77.075204} developed a TDDFT version of SBE, which formulate TDDFT in terms of the density matrix. By replacing the Kohn-Sham Hamiltonian $H_{\rm KS}$ in Eq.~\eqref{ks} with the SBE Hamiltonian $H_{\rm SBE}$ in Eq.~\eqref{Hsbe}, then the TDSE Eq.~\eqref{TDSE} yields the TDDFT version of the SBE. More details can be found in Ref.~\cite{PhysRevB.77.075204,PhysRevB.79.233201}. 

We remark that the experimental band gap of ZnO is $E_{\rm g}$ = 3.30 eV~\cite{zno3.3ev}, while our DFT+U value is 2.81\,eV.
To maintain consistency in the comparison of the different theoretical approaches, we decided to use the value of 2.81 eV for the calculations using SBE and the Keldysh model. Similarly, the volume employed is always the one obtained from DFT relaxation.
The effects of the bandgap discrepancy are further discussed in Section S3 of the SM. 

\section{Results and discussion}

In our analysis, we examine various pump laser wavelength scenarios and compare the number of excitation electrons calculated with all three models, $\rm n^{Keldysh}_{ex}$,  $\rm n^{SBE}_{ex}$,  and $\rm n^{TDDFT}_{ex}$, with the threshold electron density in the conduction band $n_{\rm th}$ for lasing. The $n_{\rm th}$ value is slightly affected by the geometry of the nanowire, but for large diameters, it quickly approaches a fixed value, which is determined solely by the semiconductor active medium.  
We use the reference value $n_{\rm th} =2 \times 10^{25} m^{-3}$, suggested for bulk ZnO in Ref.~\cite{Hollinger2,nth}. Here  $\rm n^{Keldysh}_{ex} = \rm W^{Keldysh}_{ex} \cdot T$, with  pulse duration T of 12.1 fs for the $0.8 \ \mu$m case and 31.5 fs for 3.5 $\mu$m case.
Because Eq.~\ref{ntddft} is in velocity gauge, we define the calculated excitation electron number as the stable read-out value after the end of the pumping pulse for SBE and TDDFT, when no external vector potential is present anymore.

\subsection{Electron dynamics evolution}

\begin{figure}[ht!]
\centering
\includegraphics[width = 0.9\columnwidth]{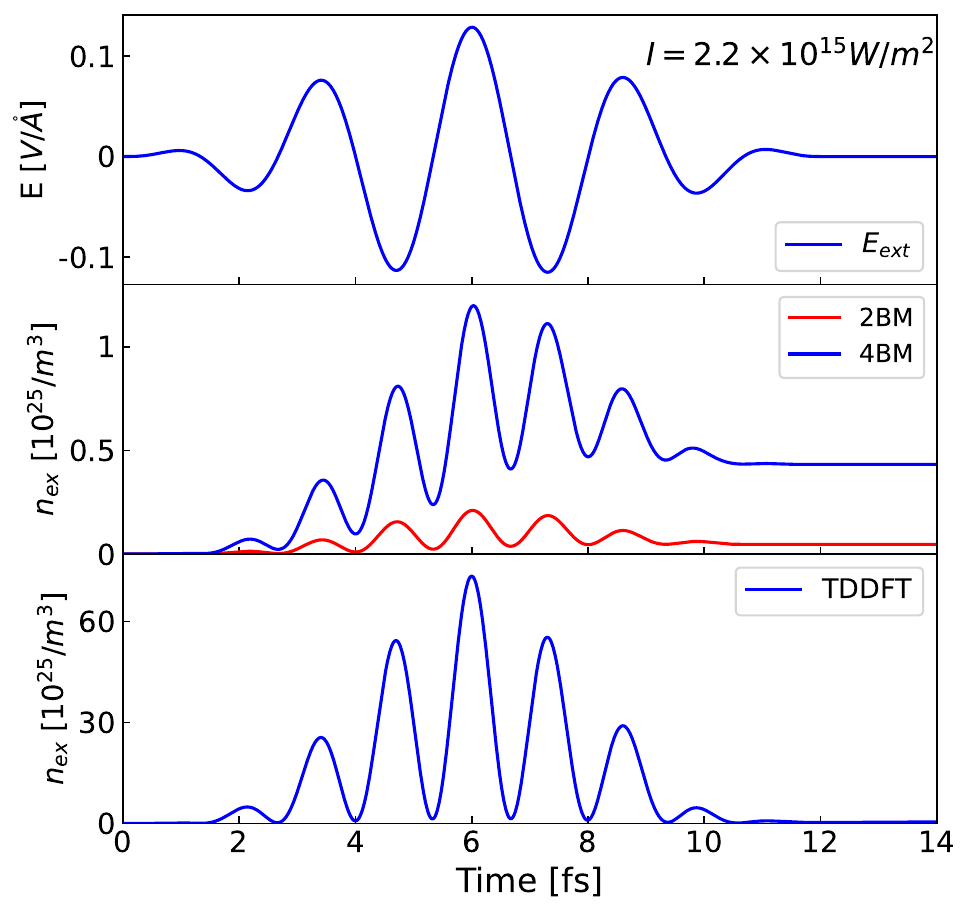}
\caption{Electron dynamics evolution of ZnO. The top panel shows the external pump laser field $\boldsymbol{E}_{{\rm ext}}$. The middle panel displays the evolution of the excited electron density $n_{\rm ex}(t)$ according to the SBE, with a 4BM (blue) and a 2BM (red). The bottom panel illustrates the evolution of $n_{\rm ex}(t)$ in TDDFT. 
The pump laser intensity is $2.2 \times 10^{15} W/m^2$.} 
\label{figcompare} 
\end{figure}

In Fig.~\ref{figcompare}, we show the evolution of the density of excited electrons $n_{\rm ex}$, in response to the external pump laser. The upper panel illustrates the near-IR external electric field, $\boldsymbol{E}_{\rm ext}$, with a wavelength $\lambda = 0.8 \ \mu$m, a photon energy $\hbar \omega$ = 1.55 eV, and a field strength $E_0 = 0.13 \ \text{V/}\mathring{A}$, which corresponds to an intensity of $2.2 \times 10^{15} \ \text{W/m}^2$. The middle panel shows the evolution of the excited electron density $n_{\rm ex}$ obtained from the SBE , with the 4BM depicted in blue and the 2BM in red. The bottom panel presents the evolution of $n_{\rm ex}$ for TDDFT. The stable read-out values of $n_{\rm ex}$ after the pulse for each method are listed in Table \ref{5E15}. 

Notably, the $n_{\rm ex}$ obtained from the 2BM approach is an order of magnitude lower than that from the other two methods. 
One reason for this discrepancy is due to broken bands degeneracies (see Fig.~\ref{band}) and the consequent loss of oscillator strength for the transitions along high symmetry lines.
Another important factor is the inherent limitation of a 2BM, which is often insufficient to fully describe electron dynamics.
Similar conclusions have been drawn for high-harmonic generation in solids at comparable laser intensities, where up to 51 bands needed to be included for accurate results using the SBE~\cite{HHGSBE50}.

\begin{center}
\begin{table}[ht!]
\begin{tabular}{|l|c|}
\hline  
Method &   $n_{\rm ex} [1/m^3]$  \\
\hline  
2BM & 4.61 $\times 10^{23}$ \\
4BM  & 4.32 $\times 10^{24}$\\
TDDFT  & 2.94 $\times 10^{24}$\\
\hline
\end{tabular}
\caption{Density of excited carriers at the end of the laser pulse from the different models, compared to the experimental data.}
\label{5E15}
\end{table}
\end{center}

\begin{figure}[ht!]
\centering
\includegraphics[width = 0.9\columnwidth]{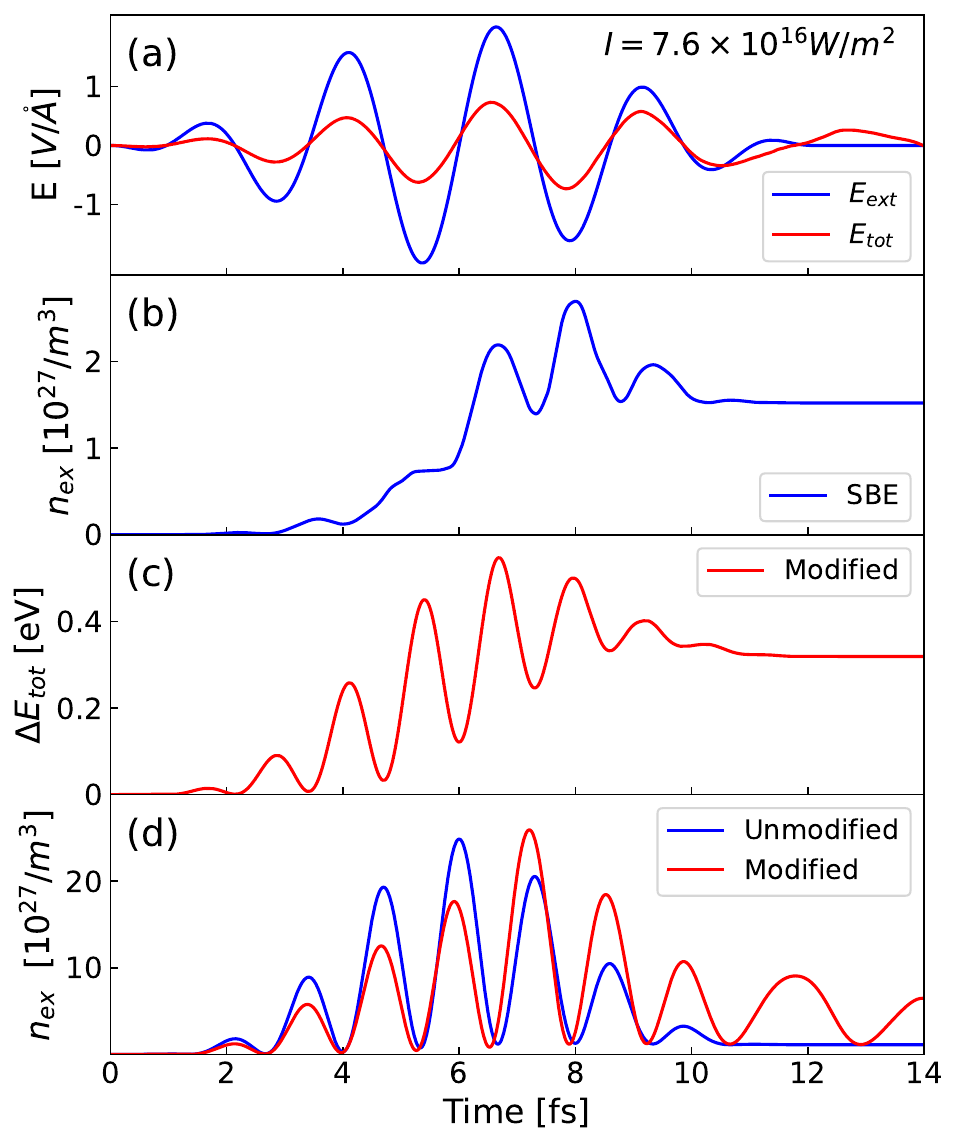}
\caption{Electron dynamics evolution of ZnO. 
(a) shows the external electric field $\boldsymbol{E}_{{\rm ext}}$ (blue), and the total electric field $\boldsymbol{E}_{\rm tot}$ (red) of TDDFT. 
(b) displays the evolution of the excited electron density $n_{\rm ex}(t)$ for SBE with 4BM. 
(c)  illustrates  the change in total energy $\Delta E_{\rm tot}$ for the ``modified" case of TDDFT. 
(d) displays  the evolution of $n_{\rm ex}(t)$ for the ``modified" case (red) and the ``unmodified" case (blue) of TDDFT, details in script.
The pump intensity is $\rm 7.6 \times 10^{16} W/m^2$.} 
\label{HighIntensity} 
\end{figure}

For higher intensities, numerical instabilities become significant in the SBE calculation. In Fig.~\ref{HighIntensity}, the external laser intensity is $\rm 7.6 \times 10^{16} \ W/m^2$. Panel (b) displays the evolution of the excited electron density $n_{\rm ex}(t)$ for SBE with the 4BM. Under this intensity, electrons tend to transition to higher energy bands, and even the 4BM is insufficient to fully describe the electron behavior, leading to some discordant parts in the evolution curves. Therefore, in the following, we restrict ourselves to  pump intensity  for SBE below $2 \times 10^{16} \ W/m^2$.
 
Before discussing the TDDFT data, the definition of the laser intensity needs to be clarified. The pulse intensity mentioned throughout the paper is the intensity measured outside the crystal, as per experimental convention. However, the corresponding intensity in the crystal during the TDDFT simulation requires careful treatment.

The modification has to be consistent with the vector potential included in Eq.~\eqref{ks}. If, in the first term $\frac{1}{2m_e} (\boldsymbol{p} + \frac{e}{c} \boldsymbol{A}_{\rm tot}(t))^2$ on the right side of Eq.~\eqref{ks}, we substitute $\boldsymbol{A}_{\rm ext}(t)$ for $\boldsymbol{A}_{\rm tot}(t)$, the equation remains approximately valid provided that the induced-field propagation is suppressed to avoid double counting. In this scenario, the laser intensity used is the experimental value outside the crystal, corresponding to the vector field $\boldsymbol{A}_{\rm ext}(t)$. We denote this scenario as the ``unmodified" case, as we are not modifying the laser intensity. This approach, for example, was utilized for comparative analysis with the Keldysh model for crystalline silicon in Ref.~\cite{Keldysh1}.

When employing $\boldsymbol{A}_{\rm tot}(t)$ in Eq.~\eqref{ks} for the dynamic calculation, we should consider that the total field that the electrons experience inside ZnO is reduced compared to the external field $\boldsymbol{A}_{\rm ext}(t)$ due to reflection losses. The electric field reduction can be obtained from the Frenels coefficients, dividing by the real part of the dielectric function $\epsilon_1$ of ZnO, following, e.g., Ref.~\cite{NiO,E_ind}. The complex dielectric tensor $\epsilon$ can be  calculated in linear response using Eq.~\eqref{eps}, more details are given in section III.D. 
In the panel (a) of Fig.~\ref{HighIntensity}, the external laser field $\boldsymbol{E}_{\rm ext}$ (blue),  obtained using Eq.~\eqref{EA}, has a strength of $2.007 \ V/\mathring{A}$, while the total laser field  $\boldsymbol{E}_{\rm tot}$ (red), obtained by adding the induced electric field to the external field, has a strength of $ 0.734 \ V/\mathring{A}$. For comparison, $\boldsymbol{E}_{\rm ext} / \epsilon_1 = 0.741 V/\mathring{A}$ with a calculated $\epsilon_1 = 2.710$. The values of $\boldsymbol{E}_{\rm tot}$ and $\boldsymbol{E}_{\rm ext} / \epsilon_1$ are nearly the same, with a mere 0.9\% difference. Some references (for example Ref.~\cite{E_ind}) use the calculated $\boldsymbol{E}_{\rm tot}$, while others (for example Ref.~\cite{NiO}) use rather $\boldsymbol{E}_{\rm ext} / \epsilon_1$. We use here directly $\boldsymbol{E}_{\rm tot}$  in the following calculations and define this approach as ``modified'', considering that we are modifying the external pulse by adding the induced reponse of the medium.

In summary, we can calculate the inner-crystal intensity $I_{in}$ to be $7.15 \times 10^{16} W/m^2$ with $\boldsymbol{E}_{\rm tot}$. 
The outer-crystal intensity $I_{out}$
can be calculated from  $I_{in}$ by $I_{out} = I_{in} / (1-R)$, where the reflectivity $R$ of normal incidence is given in  Eq.~\eqref{reflection} with complex refractive index $N = n+i\kappa$, which can be  calculated from the complex dielectric function $\epsilon = \epsilon_1 + i \epsilon_2$ in Eq.~\eqref{neps},

\begin{equation}
R = \left|\frac{n+i\kappa-1}{n+i\kappa+1}\right|^2.
\label{reflection}
\end{equation}

\begin{equation}
n = \sqrt{ \frac{(|\epsilon| + \epsilon_1)}{2}} ,\ \ \ \  
\kappa = \sqrt{\frac{(|\epsilon| - \epsilon_1)}{2}}.
\label{neps}
\end{equation}

In the case of ZnO, in Fig.~\ref{HighIntensity}, the calculated $n$ = 1.647, $\kappa$ = 0.039, and the reflectivity $R$ is 6\%. Therefore, the outer-crystal intensity $I_{out}$  is  $7.6 \times 10^{16} W/m^2$ as indicated in the panel (a) of Fig.~\ref{HighIntensity}. Due to the above described modification process, we refer to this case as ``modified".  The TDDFT results of the bottom panel in Fig.~\ref{figcompare}  are also obtained with this ``modified" procedure and the outer-crystal intensity is $2.2 \times 10^{15} W/m^2$.

Panel (c) of Fig.~\ref{HighIntensity} shows the evolution of the total energy difference $\Delta E_{\rm tot}(t)$ in eV for the ``modified" case. 
The evolution of the excited electron number $n_{\rm ex}(t)$ for the ``modified" case (red) is shown in the panel (d) of Fig.~\ref{HighIntensity}.  The applied $\boldsymbol{E}_{\rm ext}$ field ends at 12.1 fs, but the corresponding $\boldsymbol{E}_{\rm tot}$ and $n_{\rm ex}$ still exhibit oscillations until the end of the pulse at 14 fs. 
Ref.~\cite{oscillation} attributes this behavior to plasma oscillations of the excited electrons in the conduction bands. 
Indeed, as no dissipation mechanism is included, the total energy is conserved, but transfer of energy will occur between the electrons and the Maxwell field, leading to oscillations in the induced vector potential.

To correct this, one can employ a gauge-dependent definition of the ground-state orbitals as in Ref.~\cite{oscillation}.  We assume that the gauge field $\mathbf{A}(t)$ is slowly varying and spatially uniform. Then, the  ground-state orbitals can be adiabatically evolved by replacing the Bloch wave number $\mathbf{k}$ with $\mathbf{k} + \mathbf{A}(t)$ and replacing the Bloch wave functions $\phi_{n\mathbf{k}}$ with $\phi_{n\mathbf{k}+\mathbf{A}(t)}$. 
Since the entire $\mathbf{k}$ region is occupied in the dielectrics, this adiabatic evolution does not produce any real excitation. Consequently, compared to $n^{\rm TDDFT}_{\rm ex}(t)$ in Eq.~\eqref{ntddft}, the corrected normalized density of excited electrons $n^{\rm TDDFT}_{\rm ad}(t)$ can be calculated with respect to the adiabatically-evolved  orbitals, 

\begin{equation}
n^{\rm TDDFT}_{\rm ad}(t) = \frac{2}{V} \sum _{nn\prime \mathbf{k}} w_{\mathbf{k}} (\delta_{nn^{\prime}} - | \langle \phi_{n\mathbf{k}+\mathbf{A}(t)}|\psi_{n^{\prime}\mathbf{k}}(t) \rangle|^2) \,.
\label{ntddftad}
\end{equation}

The ``modified" result can therefore be obtained by connecting the minima of the oscillations~\cite{oscillation} of the red curve in the bottom panel of Fig.~\ref{HighIntensity}. This enables us to read out the stable value of $n_{\rm ex} = 1.146 \times 10^{27}/m^3$ using the minimum points after the pulse ends for the ``modified" case. 

We can now compare with the results obtained in the ``unmodified" case. This approach has the additional advantage that it does not include the computational cost of the induced vector potential and is therefore faster by a factor of two than calculations in the ``modified" case.
The evolution of $n_{\rm ex}(t)$ for the ``unmodified" case is depicted by the blue curve in panel (d) of Fig.~\ref{HighIntensity}. We can observe a perfectly flat line after the pulse ends at 12.1 fs, with a final value of $n_{\rm ex} = 1.126 \times 10^{27}/m^3$. 

This value is very close to the one obtained in the ``modified'' calculation ($1.7\%$ difference) at a reduced computational cost.  The $n_{\rm ex}$ of the ``unmodified" case is slightly lower than the one calculated in the ``modified" case. This trend is also observed for an intensity of $5\times 10^{15} W/m^2$, as reported in Section S4 of the SM.

To strike a balance between calculation cost and accuracy, we employ the different calculation schemes, ``modified" and ``unmodified", for the shorter near-IR 0.8  $\mu$m and longer mid-IR 3.5 $\mu$m cases, respectively.

\subsection{Near-infrared pump laser}

\begin{figure}[ht!]
\centering
\includegraphics[width = 0.9\columnwidth]{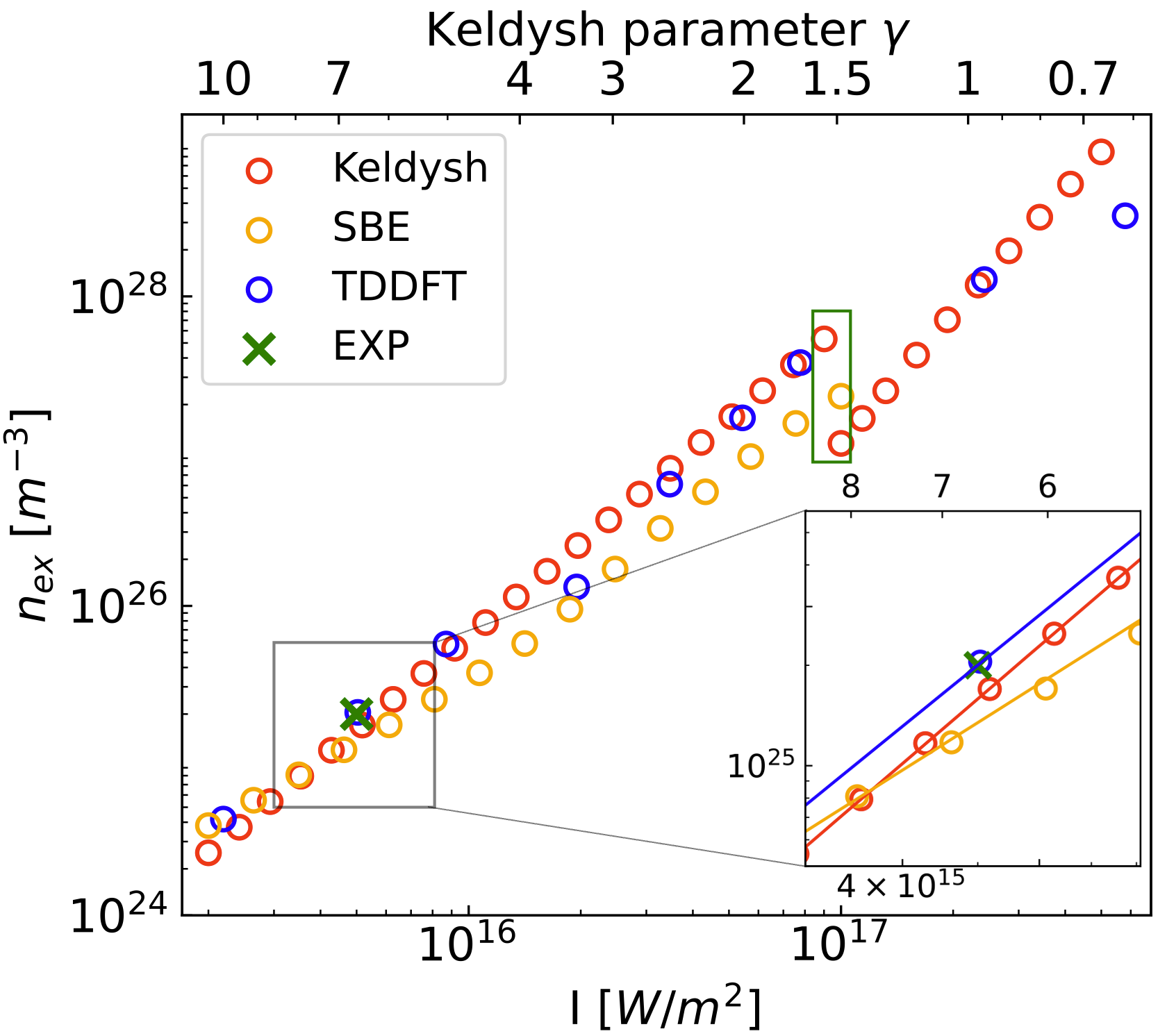}
\caption{The density of excited electrons $n_{\rm ex}$ for different pump laser intensity I.  The data points are Keldysh (red), SBE (orange), TDDFT (blue), and experimental data (green cross)~\cite{Hollinger2,nth}. The Keldysh parameter $\gamma$ is recorded on the upper x-axis. The  near-IR pump laser has a wavelength $\lambda = 0.8 \ \mu$m corresponding to a photon energy of $\hslash \omega$ = 1.55 eV.}  
\label{zno_800} 
\end{figure}

In Fig.~\ref{zno_800}, we compare the calculated excited electron number per unit volume $n_{\rm ex}$ in $m^{-3}$ for different pump intensities considering all three calculation methods: Keldysh (red), SBE (orange), TDDFT (blue, ``modified'').  
The pump is a near-IR laser with a wavelength $\lambda = 0.8 \ \mu$m and a photon energy $\hslash\omega$ = 1.55 eV. 
The laser intensity ranges from $2\times10^{15}$ to $6\times10^{17}$ W/m$^2$. 
The green cross represents the experimental pump-laser threshold intensity $I_{\rm th}^{0.8}$ = $5\times10^{15}$ W/m$^2$~\cite{Hollinger2}, and the  reference carrier density  $n_{\rm th} = 2 \times 10^{25}/m^{3}$~\cite{Hollinger2,nth}. The Keldysh parameter $\gamma$ is recorded on the upper x-axis. 

In the Keldysh calculation points (red circles), a distinct discontinuity is highlighted by a green rectangular frame. This region encompasses two data points corresponding to pump intensities $I = 10^{17}$ and $1.06 \times 10^{17} W/m^2$, with values of $(\frac{\tilde{E}_g}{\hbar \omega} + 1)$ being 2.99 and 3.00, respectively. As  $k = \lfloor \frac{\tilde{E}_g}{\hbar \omega} + 1 \rfloor$ in Eq.~\eqref{eq:k} indicates the $k$-photon absorption process, these data points correspond to the transition from two-photon absorption (2PA) to three-photon absorption (3PA) regimes in the Keldysh model, respectively, leading to the observed discontinuity. 

In the double-log coordinate system, the linear relation signifies that the excited electron density $n_{\rm ex}$ is proportional to the $m$-th power of pump intensity $I_{\rm p}$: 
$ \log(n_{\rm ex}) \propto \log(I_{\rm p}^m) \propto m \cdot \log(I_{\rm p})$, 
where $m$ denotes the slope. The value of $m$ does not necessarily coincide with the $k$ of Keldysh model.

For the initial 21 Keldysh points, at lower intensities,  $k = m = 2$, indicating the predominance of a 2PA mechanism. However, for the last 10 Keldysh points, at higher intensities, $k =3$, $m = 2.7$, and the corresponding Keldysh parameter $\gamma$ varies from 1.5 to 0.7. This indicates that the absorption mechanism above an intensity of $10^{17}$ W/m$^2$ is not only due to 3PA but also involves tunneling ionization. 

The inset  of Fig.~\ref{zno_800} provides a magnified view of the calculation points in a specific region of interest, near the experimental threshold intensity $I_{\rm th}^{0.8}$. 
For each calculation method, we  performed a linear fitting of the low-intensity points (specifically, 14 for Keldysh,  3 for TDDFT, and 10 for SBE). This procedure allows us to accurately extract by interpolation the $n_{\rm ex}$ value at $I_{\rm th}^{0.8}$ = 5$\times10^{15}$ W/m$^2$. The slope $m$ of the linear fitting curve,   the corresponding density of excited electrons $n_{\rm ex}$ at $I_{\rm th}^{0.8}$ and relative error
are reported in Table.\ref{t1}. 

\begin{center}
\begin{table}[ht!]
\begin{tabular}{|l|c|c|c|}
\hline 
Method & $m$ & $n_{\rm ex} [m^{-3}  ]$  & Rel. Error \\
\hline 
Exp.~\cite{Hollinger2,nth}& - & $2 \times 10^{25}$ & -\\
Keldysh & 2.00 &$1.6 \times 10^{25}$ & $20.7\%$\\
SBE & 1.47 & $1.4\times 10^{25}$ & $32.7\%$ \\
TDDFT & 1.89 &$2.0\times 10^{25}$& $0.1\%$\\
\hline
\end{tabular}
\caption{Comparison of the slope $m$ and excited-electron density $n_{\rm ex}$ according to the different calculation methods for $\lambda=0.8 \ \mu m$.}
\label{t1}
\end{table}
\end{center}

For the near-IR pulse ($\lambda$=0.8 $\mu$m), in the vicinity of $I_{\rm th}^{0.8}$, the Keldysh model exhibits a slope $m$ exactly equal to 2, clearly indicating a 2PA mechanism. In contrast, TDDFT and SBE have slopes $m$ of 1.89 and 1.47, respectively, hence smaller than 2. This suggests that the excitation mechanism is 2PA-dominated but also involves some degree of tunneling assistance. Such differences between the models are expected as the Keldysh model simplifies the band structure to two energy levels and is no more valid at very high field intensities.

If we compare calculations and experiment, the density of excited electrons $n_{\rm ex}$ calculated using TDDFT shows the best agreement with experimental data.
The small error in SBE arises from the limitation of the 4BM. TDDFT allows for more electrons to transition to higher energy levels, resulting in a larger slope $m$ and higher $n_{\rm ex}$ than SBE. The dephasing time $T_2$ of SBE also affects the results, as discussed in Section S2 of the SM.

\subsection{Mid-infrared pump laser}

\begin{figure}[ht!]
\centering
\includegraphics[width = 0.9\columnwidth]{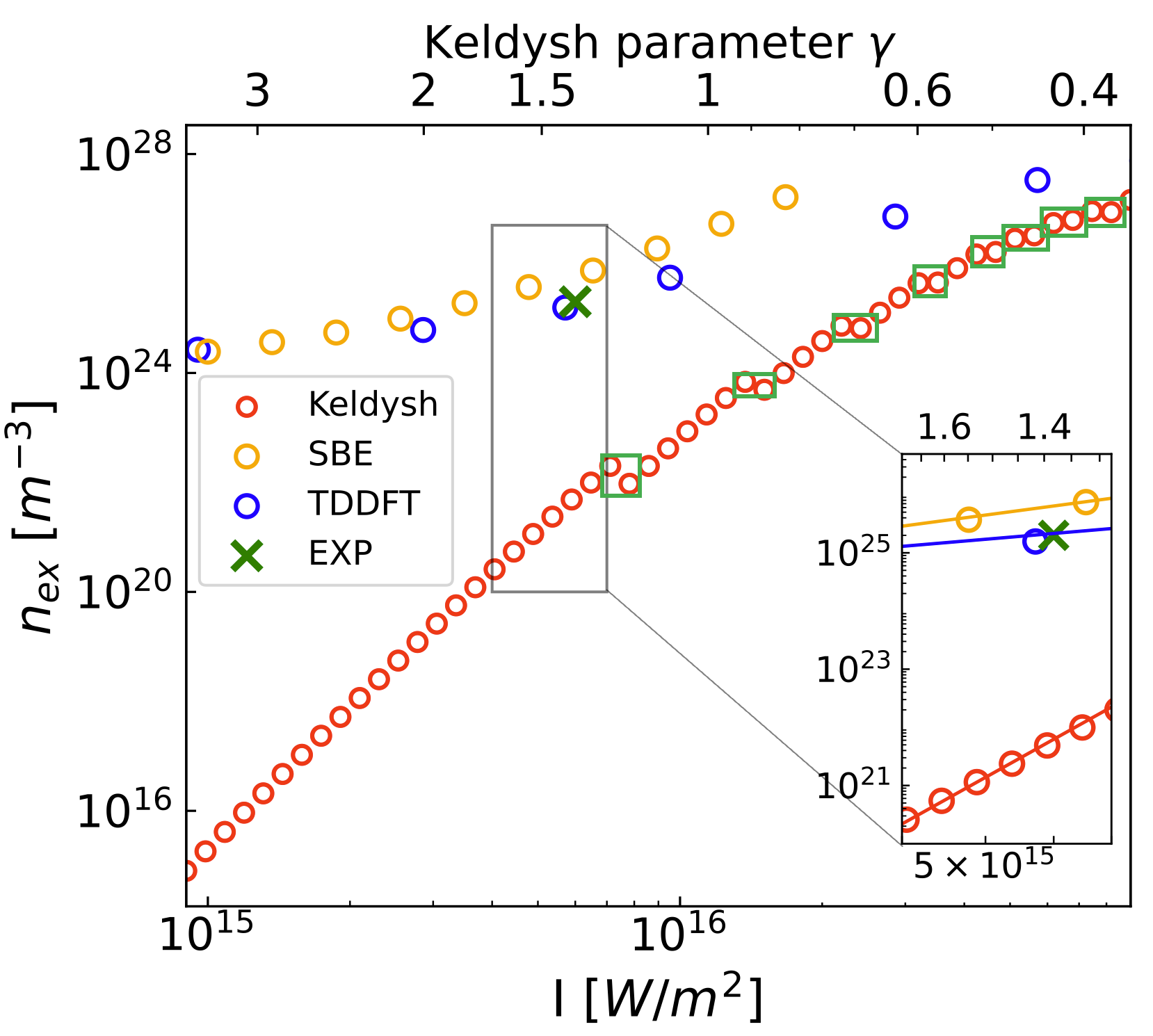}
\caption{Same as Fig.~\ref{zno_800}, but for a mid-IR pump laser with a wavelength $\lambda = 3.5 \ \mu$m corresponding to a photon energy of $\hslash \omega$ = 0.354 eV. }  
\label{zno_3500} 
\end{figure}

Now, we turn to the mid-IR case. The pump laser has a wavelength $\lambda = 3.5 \ \mu m$ and a corresponding photon energy $\hslash \omega$ = 0.354 eV.  
In Fig.~\ref{zno_3500}, we compare the excited electron density $n_{\rm ex}$(per $m^{3}$) across different pump intensities for all calculation methods: Keldysh (red), SBE (orange), TDDFT (blue; ``unmodified"), and reference data (green cross)~\cite{Hollinger2, nth}. The Keldysh parameter $\gamma$ is shown on the upper x-axis. 

In the Keldysh data, discontinuous regions are highlighted by green rectangular frames. They indicate changes in the parameter  $k$ of Eq.~\eqref{eq:k}.  The lowest frame illustrates the conversion of $k$ from 9PA to 10PA as the intensity increases from  $7.12 \times 10^{15}$ to $7.82 \times 10^{15}$ W/m$^2$.  Subsequent frames show the conversions of $k$ from 10 to 11, 12, ... , 17, indicating transitions to higher multi-photon processes. 

The inset of Fig.~\ref{zno_3500} provides a magnified view of the calculation points  near the experimental threshold intensity $I_{\rm th}^{3.5}$. 
As for the near-infrared laser, we fit the low-intensity points to a linear curve (23 for Keldysh,  4 for TDDFT, and 8 for SBE).
The slopes $m$ obtained from the linear fitting, the corresponding density of excited electrons $n_{\rm ex}$ at $I_{\rm th}^{3.5}$ and the relative difference are tabulated in Table.\ref{t2}.

\begin{table}[ht!]
\begin{tabular}{|l|c|c|c|}
\hline 
Method&  $m$ & $n_{\rm ex} [m^{-3}]$  & Rel. Error \\
\hline  
Exp. ~\cite{Hollinger2, nth} & - & $2 \times 10^{25}$ & - \\
Keldysh & 8.27 & $6.2 \times 10^{21}$ &  - \\
SBE & 1.97 &  $6.4 \times 10^{25}$& $219\%$\\
TDDFT & 1.26 & $2.2 \times 10^{25}$ & $7.7\%$ \\
\hline
\end{tabular}
\caption{Comparison of the slope $m$ and excited-electron density $n_{\rm ex}$ according to the different calculation methods for $\lambda=3.5 \ \mu m$. }
\label{t2}
\end{table}

For the Keldysh calculations, a linear fit of the initial 23 points yields a slope $m$ of 8.27. However, using only the first 10 points for linear fitting increases $m$ to 8.60. This indicates that the slope $m$ decreases as the intensity increases within an intensity interval with the same value of $k$. 
These initial 23 data points from the Keldysh dataset all fall within the 9-photon absorption regime with $k = 9$. The slope $m =8.27 < k$, suggests that, even in the Keldysh model, the absorption mechanism includes some degree of tunneling. The value of $n^{\rm Keldysh}_{\rm ex}$ is significantly lower than the reference data $n_{\rm th}$, which is consistent with Ref.~\cite{Keldysh1}. Thus MPA alone cannot provide a sufficient number of excited electrons to achieve population inversion and lasing.

The SBE calculation exhibits a 3-times higher $n_{\rm ex}$ than the experimental data. For the mid-infrared pulse, the simulated propagation time is significantly longer than in the previous calculation with $\lambda =$ 0.8 $\mu$m, providing electrons with more opportunities to transition to higher energy levels. The limitations of the 4BM become more pronounced, and propagation instability becomes substantial, leading to inaccurate predictions. However, SBE performs much better than the Keldysh model. 

The TDDFT calculations provide the best value of $n_{\rm ex}$ when compared to the experimental data. However, also the TDDFT value is $7.7\%$ larger than experiment. This could be due to missing decoherence effects, which become more important for large propagation time.

One possible cause of disagreement between calculations and experiment can come from the difference between the calculated DFT+U bandgap of 2.81 eV and the experimental value of 3.3 eV. The higher bandgap makes the electron transitions more challenging, resulting in a lower $\rm n_{ex}$. We performed Keldysh and SBE calculations using the experimental band gap and observed modifications of the excitated-electron densities, in particular for the Keldysh model, as discussed in S3 of the SM. We conclude that the TDDFT results could be further improved by employing a more accurate exchange-correlation functional or other methods to obtain a more precise bandgap value. The choice of a different exchange-correlation functional could also have impact in the propagation of the Kohn-Sham equations. 

The TDDFT deviations of $0.1\%$ and $7.7\%$ for 0.8 $\mu$m and 3.5 $\mu$m, respectively, suggest that the ``modified" propagation scheme for the total gauge field $\boldsymbol{A}_{\rm tot}(t)$ is more accurate,  however it is also significantly more expensive. 
Overall, we conclude that the agreement is already extremely good.

The slopes $m$ of the TDDFT and SBE curves are 1.26 and 1.97, respectively, suggesting that the transition mechanism is predominantly tunneling. 
In this scenario, under the influence of the strong laser field, the bands bend significantly, leading to a reduction in the potential barrier between the maximum of the valence band and the minimum of the conduction band. Electrons can then directly transition through this lowered barrier via the quantum tunneling effect, even without sufficient energy to overcome the band gap. 
The electronic excitations are determined by the interplay of the dominant intraband tunneling and the auxiliary interband MPA process.   
As the pump intensity continues to increase, even Keldysh shifts towards to tunneling domination ($\gamma < 1$), resulting in a reduced difference between the TDDFT results and the Keldysh model.

\subsection{Dynamical Franz-Keldysh Effect}

Under irradiation by an intense laser field, the dielectric function $\epsilon(\omega)$ undergoes transient modification. The modification caused by a static electric field is known as the Franz-Keldysh effect (FKE)~\cite{franzFKE, keldyshFKE}, while the modification caused by an alternating electric field is known as the Dynamical Franz-Keldysh effect (DFKE)~\cite{DFKE1996, DFKE1998}. 

The Keldysh parameter $\gamma$ can also distinguish between these two regimes. The MPA applies for $\gamma \gg 1$,  and the static FKE is appropriate for $\gamma \ll 1$, in the tunneling regime.   For  $\gamma \approx 1$, the DFKE phenomena is expected.  We choose the experimental mid-IR pulse with $I_{\rm th}^{3.5} = 6\times 10^{15}W/m^2$, $\lambda = 3.5 \ \mu$m, and $\gamma = 1.38$ to explore DFKE in ZnO, as the other experimental near-IR laser has a much larger $\gamma = 6.63$ which is too large for our purposes.

We apply a full-frequency $\delta$-function $E_0\delta(t-t_0) \hat{e}_z$  to ``kick" the electrons in our simulation cell and probe the optical information after electrons are excited to conduction bands by the experimental mid-infrared pump pulse. Here, $t_0$ = 29 fs is the time delay between the probe and pump pulse, and the total propagation time is 36.3 fs.  The ``kick" strength is $A_0$ = 2.7 eV after gauge transform via Eq.~\eqref{EA}, and  $\hat{e}_z$ is the polarization direction of the ``kick". 

\begin{figure}[ht!]
\centering
\includegraphics[width = 0.9\columnwidth]{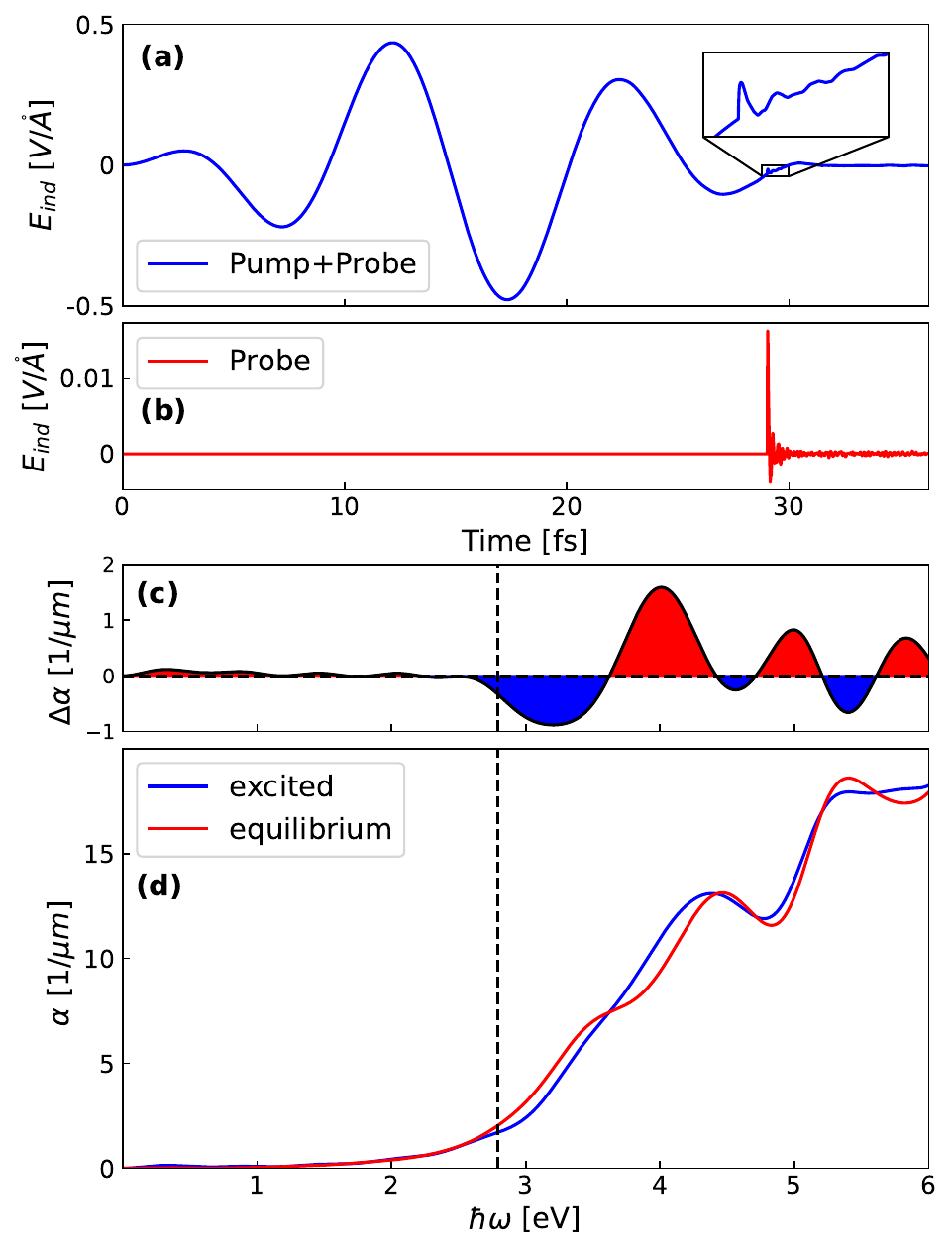}
\caption{(a) The induced electric field $\boldsymbol{E}_{\rm ind}$ of the mid-IR $pump$ pulse and weak ``kick" $probe$ with 29 fs delay. (b) The extracted ``kick" $probe$ response is equal to  $(pump+probe)$ - $pump$.  (c) The difference of absorption coefficients $\Delta \alpha(\omega)$ calculated from (d) the absorption coefficient in the excited and equilibrium states. The black dashed  line represents the calculated DFT+U bandgap value 2.81 eV.}  
\label{dfke} 
\end{figure}

The induced-electric field evolution due to  the pump pulse and ``kick" probe are illustrated in panel (a) of Fig.~\ref{dfke}. The amplitude of the pump pulse is 1000 times larger than the one of the probe, hence, near 29 fs, the ``kick" is observed as a minuscule bump, that is magnified for clarity in the inset of Fig.~\ref{dfke}.

Next, we carried out a pure pump propagation and calculated the difference of the induced vector potential $\mathbf{A}_{\rm ind}$ between ``pump+probe'' and ``pump only'' simulations to extract the response to the ``kick'' after the electrons are excited to the conduction bands by the pump pulse. This result is shown  in the panel (b) of Fig.~\ref{dfke}. For consistency, we convert the vector potential $\mathbf{A}$ to the electric field $\mathbf{E}$ via Eq.~\eqref{EA}.

From the information of induced field  $\mathbf{A}_{\rm ind}(t)$, we can calculate the dielectric function $\epsilon(\omega)$ as given by  Eq.~\eqref{eps}. Here, the parameters  $t_0$ and $A_0$ represent the delay time and strength of the probe pulse, respectively. $\omega$ is the laser frequency for $3.5 \ \mu m$ case, and the time step $d t  =  3.6 \times 10^{-4}$ fs.  The denominator is the degenerate form of $\int^{T}_{0_+} e^{i\omega t} \frac{d\mathbf{A}_{\rm ext}(t)}{dt} dt $ with $\mathbf{E}_{\rm ext} =- 1/c \cdot  d \mathbf{A}_{\rm ext}(t)/{dt}\ = E_0\delta(t-t_0) \hat{e_z}$.

\begin{equation}
\frac{1}{\epsilon(\omega)} =  1 + \frac{\int^{T}_{0_+} e^{i\omega t} \cdot D(t) \cdot \frac{d\mathbf{A}_{\rm ind}(t)}{dt} dt}{ e^{i \omega t_0} A_0} \,.
\label{eps}
\end{equation}

The damping function $D(t)$ used to reduce the tail fluctuation during Fourier transform, and $\eta$ = 0.3 eV is the damping strength.

\begin{equation}
D(t) = \begin{cases}
       \exp (- \eta (t - t_0)), &  if \ t_0 < t < T \\
       1,  & if \  0 < t \le t_0 \\
    \end{cases}
\end{equation}

Subsequently, we can calculate the absorption coefficient $\alpha(\omega)$ using  Eq.~\eqref{eq_absorption_eps} with unit  of $1/\mu m$. 

\begin{equation}
    \alpha(\omega) = \frac{4 \pi}{\lambda}   \sqrt{ \frac{|\epsilon| - \epsilon_1}{2}} \,.
\label{eq_absorption_eps}
\end{equation}

Here, $|\epsilon|$ and $\epsilon_1$ denote the absolute value and  the real part of the dielectric function, respectively.  The absorption coefficient $\alpha(\omega)$ is utilized in the Beer–Lambert absorption law $I_{\rm out} = I_{\rm in} \cdot e^{- \alpha(\omega) d }$, where $I_{\rm in}$ and $I_{\rm out}$ represent the input and output intensity, respectively, and $d$ in $\mu$m is the absorption length. 

In panel (d) of Fig.~\ref{dfke},  we compare the absorption spectrum $\alpha(\omega)$ of excited states (blue solid line),  which reveals the optical properties after electrons are excited to the conduction bands by pump pulse, and $\alpha(\omega)$ of equilibrium states (red solid line), where the electrons are still in the valence bands without pump.

Their difference is shown in panel (c) of Fig.~\ref{dfke}, $\Delta \alpha(\omega) = \alpha_{\rm excited}(\omega) - \alpha_{\rm equilibrium}(\omega)$, illustrating the absorption changes attributed to the DFKE: slight absorption occurs below the calculated DFT+U band edge 2.81 eV; near the band edge $(\hbar\omega \sim E_g)$, the first negative blue zone, shows $\alpha_{\rm excited} < \alpha_{\rm equilibrium}$, indicating a blue shift of the band edge; and oscillatory behavior is observed above the band edge. All the characteristic features of the DFKE are reproduced for the $3.5 \ \mu m$ pump at the experimental threshold intensity $I_{\rm th}^{3.5}$.

\section{Conclusions}

We employed three complementary theoretical approaches to study the excitation mechanism in ZnO bulk: the analytical Keldysh model, the 4-bands model of the semiconductor Bloch equations (SBE), and real-time time-dependent density functional theory (TDDFT) simulations. We compared our findings with experimental reference data~\cite{Hollinger2, nth} for two different pump laser wavelengths.

For the near-IR 0.8 $\mu$m pump laser, all three theoretical methods indicate that interband multi-photon absorption is the dominant excitation mechanism near the experimental laser threshold intensity~\cite{Hollinger2} $I_{\rm th}^{0.8} = 5 \times 10^{15} W/m^2$. Additionally, SBE and TDDFT reveal that intraband tunneling plays a contributing role. Among these methods, TDDFT provides the most accurate prediction of excited electron numbers compared to reference data~\cite{nth}, with only a  $0.1\%$ difference. 

For the mid-IR 3.5 $\mu$m pump laser, the Keldysh model fails to predict a sufficiently high excited electron number~\cite{nth} $n_{\rm th}= 2\times10^{25} m^{-3}$ at the experimental laser threshold intensity~\cite{Hollinger2} $I_{\rm th}^{3.5} = 6 \times 10^{15} W/m^2$, indicating that multi-photon absorption alone is insufficient to achieve lasing.  However, TDDFT and SBE provide more consistent $n_{\rm ex}$ values compared to reference data, with TDDFT showing a better fit (only $7.7\%$ difference).  In the mid-IR pump case, the excitation mechanism is determined by the interplay of dominating intraband tunneling and auxiliary interband multi-photon absorption.

We have also successfully reproduced the dynamical Franz-Keldysh effect for the mid-IR pump case with TDDFT, observed its main characteristic features. In summary, TDDFT calculations offer an accurate description of the transformation of excitation mechanisms for different pump wavelengths and different pump intensities under ultrashort laser pulse irradiation in ZnO. Further quantitative predictions and simulations based on TDDFT are expected to boost the search for novel materials and phenomena.

\subsection*{Acknowledgments} 

This research is supported by the Collaborative Research Center (CRC/SFB) 1375 ``NOA – Nonlinear Optics down to Atomic scales" of the German Research Foundation (DFG).

\bibliography{ref.bib}

\end{document}


\title{Supplemental Material: 
Real-time simulations of laser-induced electron excitations in crystalline ZnO}

\author{Xiao Chen}
\affiliation{Research Center Future Energy Materials and Systems of the Research Alliance Ruhr, Germany} 
\affiliation{Faculty of Physics and Astronomy and ICAMS, Ruhr University Bochum, Universitätstrasse 150, 44780 Bochum, Germany}
\affiliation{Institut für Festkörpertheorie und -optik, Friedrich-Schiller-Universität Jena, Max-Wien-Platz 1, 07743 Jena, Germany}
\affiliation{European Theoretical Spectroscopy Facility}
\author{Thomas Lettau}   
\affiliation{Institut für Festkörpertheorie und -optik, Friedrich-Schiller-Universität Jena, Max-Wien-Platz 1, 07743 Jena, Germany}
\author{Ulf Peschel}
\affiliation{Institut für Festkörpertheorie und -optik, Friedrich-Schiller-Universität Jena, Max-Wien-Platz 1, 07743 Jena, Germany}
\author{Nicolas Tancogne-Dejean}  
\affiliation{Max Planck Institute for Structure and Dynamics of Matter and Center for Free-Electron Laser Science, Hamburg, 22761, Germany}
\author{Silvana Botti}  
\affiliation{Research Center Future Energy Materials and Systems of the Research Alliance Ruhr, Germany} 
\affiliation{Faculty of Physics and Astronomy and ICAMS, Ruhr University Bochum, Universitätstrasse 150, 44780 Bochum, Germany}
\affiliation{Institut für Festkörpertheorie und -optik, Friedrich-Schiller-Universität Jena, Max-Wien-Platz 1, 07743 Jena, Germany}
\affiliation{European Theoretical Spectroscopy Facility}
\date{\today}
\maketitle  

\section{Dipole matrix elements}

\begin{figure*}[htbp!]
\centering
\includegraphics[width = 0.99\textwidth]{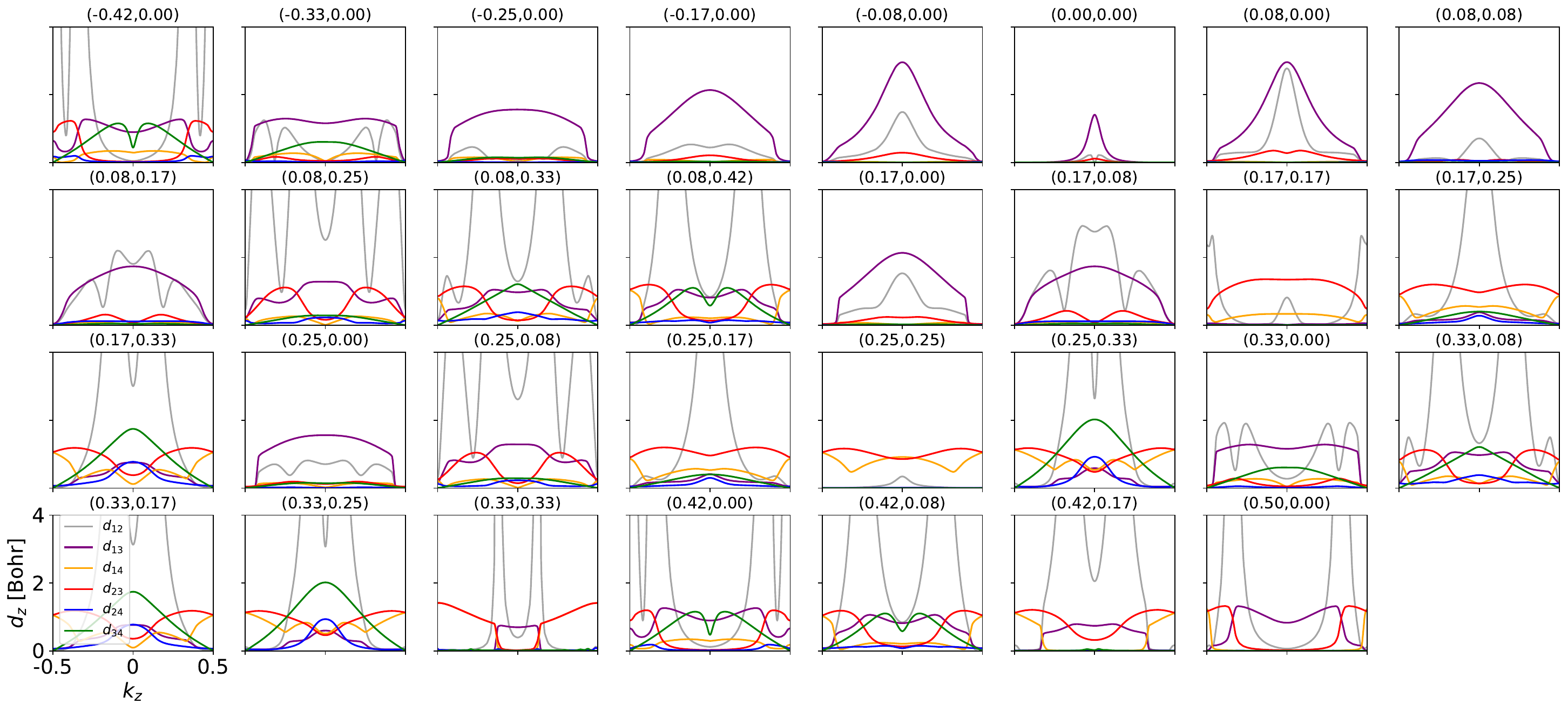}
\caption{Components along the z axis of the dipole matrix elements ($\mathbf{d}_{12}$, $\mathbf{d}_{13}$, $\mathbf{d}_{14}$, $\mathbf{d}_{23}$, $\mathbf{d}_{24}$, $\mathbf{d}_{34}$) for the ($k_x,k_y$) point with the reduced coordinates labeled in each subtitle, and as a function of $k_z$. The ranges of the axes and the color code are identical for all subfigures, with axis labels provided in the lower left panel.}
\label{dipole} 
\end{figure*}

Fig.~\ref{dipole} shows the $z$-component of the dipole matrix elements $\mathbf{d}_{z}$ used in the 4-band model (4BM) for the semiconductor Bloch equations (SBE), calculated with DFT+U.
Different colors show different matrix elements: $\mathbf{d}_{12}$, $\mathbf{d}_{13}$, $\mathbf{d}_{14}$, $\mathbf{d}_{23}$, $\mathbf{d}_{24}$ and $\mathbf{d}_{34}$.
Each subfigure presents a slice of $\mathbf{d}_{z}$ for specific ($k_x,k_y$) coordinates indicated in each subtitle. The horizontal axis represents the reduced coordinates of $k_z$.

\section{Dephasing time of SBE}

\begin{figure*}[htbp!]
\centering
\includegraphics[width = 0.9\textwidth]{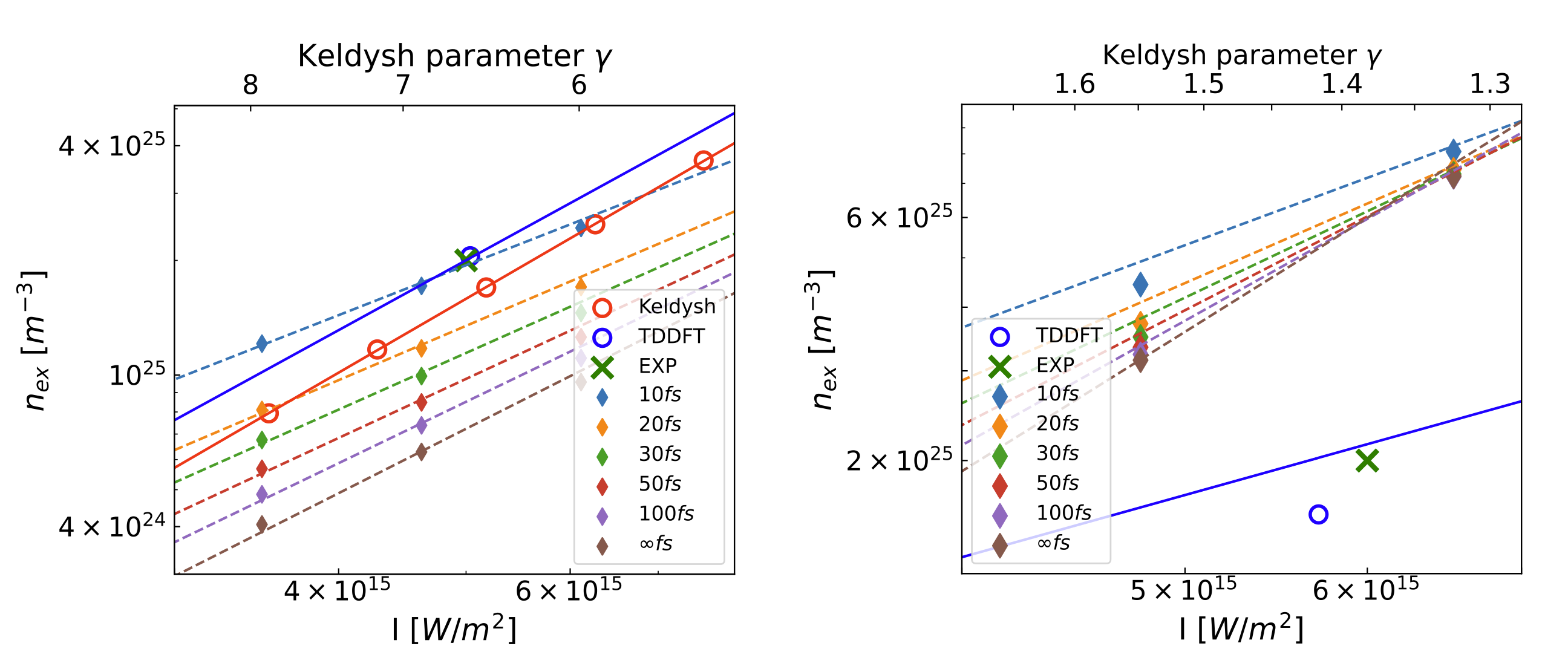}
\caption{Density of excited electrons $n_{\rm ex}$ for different dephasing times $T_2$ obtained from the SBE.  (Left) laser wavelength $\lambda = 0.8 \ \mu$m; (Right) laser wavelength $\lambda = 3.5 \ \mu$m.}
\label{diffT2}
\end{figure*}

When calculations are performed using the SBE, the density of excited electrons $n_{\rm ex}$ and the slope $m$ of the density of excited electrons with respect to the laser intensity I are significantly affected by the dephasing time $T_2$. Here, the $n_{\rm ex}$ curves obtained from the SBE for different $T_2$ values are shown in Fig.~\ref{diffT2}. The left and right panels correspond to pump-laser wavelengths of  0.8 and 3.5 $\mu$m, respectively.

In Fig.~\ref{diffT2}, the solid diamonds with different colors represent the SBE data with different $T_2$, the corresponding linear fitting curve were calculated using the initial 10 or 8 points for the 0.8 or 3.5 $\mu$m lasers, respectively. 
Due to the increased propagation instability associated with the extended propagation time for the 3.5 $\mu$m laser, and the insufficiency of the 4BM to accurately describe high-intensity scenarios, the reliable data points obtained for the 3.5 $\mu$m laser are less than those for the 0.8  $\mu$m laser.

The values of the density of excited electrons $n_{\rm ex}$ and of the slope $m$ of the curves of Fig.~\ref{diffT2} for various values of $T_2$ are documented in Table \ref{t_diffT2_800}. For both the 0.8 and 3.5 $\mu$m lasers, an increase in $T_2$ corresponds to a longer coherence time, which enhances the efficiency of the multi-photon absorption process, resulting in a gradually increasing slope $m$. 

\begin{center}
\begin{table}[ht!]
\begin{tabular}{|l|c|c|c|c|}
\hline  
$T_2$ [fs]& $m_{0.8}$ & $n^{0.8}_{\rm ex} [ 10^{25} /m^{3}  ]$  & $m_{3.5}$ &   $n^{3.5}_{\rm ex} [ 10^{25} /m^{3}  ]$ \\
\hline  
10 & 1.35 & 1.9426  & 1.67 & 7.1791 \\
20 & 1.47 & 1.3465 & 1.97 & 6.3910 \\
30 & 1.53 & 1.1427 & 2.14 & 6.1712 \\
50 & 1.60 & 0.9771  & 2.33 & 6.0296 \\
100 & 1.66 & 0.8508 & 2.52 & 5.9616 \\
$\infty$ & 1.75 & 0.7227 & 2.82 & 5.9901\\
\hline
\end{tabular}
\caption{Comparison of the slope $m$ and the density of excited electrons $n_{\rm ex}$ obtained from the 4BM-SBE with different dephasing times $T_2$ for pulses wavelengths of 0.8 $\mu$m and 3.5 $\mu$m.}
\label{t_diffT2_800}
\end{table}
\end{center}

\section{Bandgap difference}

The bandgap value plays a crucial role in electron transition calculations. In our TDDFT calculations we use a DFT+U bandstructure with a bandgap of 2.81\,eV, whereas the experimental bandgap is about 3.3\,eV. In this section, we assess the impact of this discrepancy in the value of the bandgap.

To simulate the experimental bandgap with a Keldysh model, we can easily adjust the value $E_g$ from 2.81 eV to 3.3 eV. When solving the SBE, we can add the difference 0.49 eV to the $E_2^{\mathbf{k}}$ term (for these calculations $T_2$ is set to 20 fs). 

\begin{figure*}[htbp!]
\centering
\includegraphics[width = 0.9\textwidth]{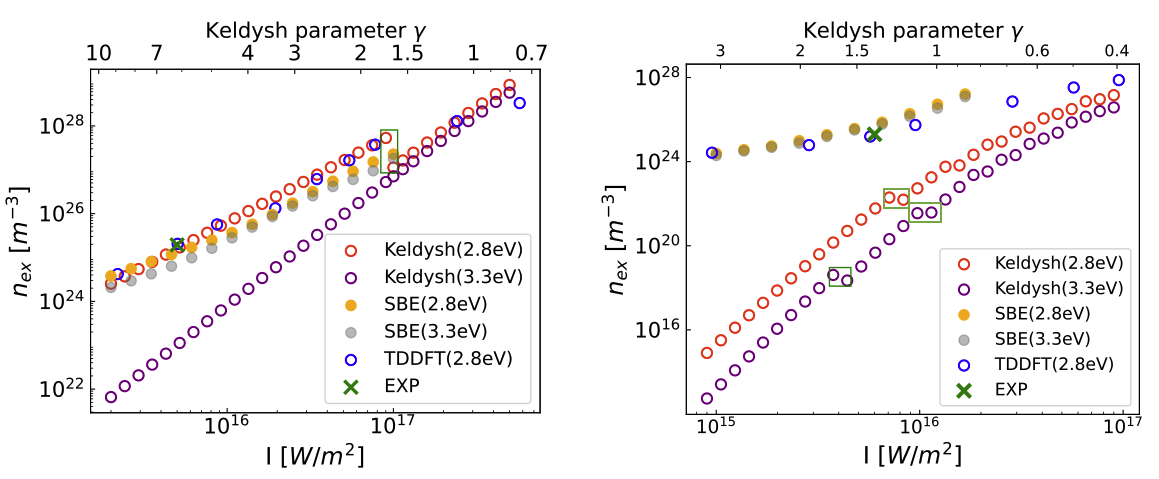}
\caption{Density of excited electrons $n_{\rm ex}$ for different bandgap values: 2.81 eV from DFT+U band structures and 3.3 eV from experiment.  (Left) $\lambda = 0.8 \ \mu$m; (Right) $\lambda = 3.5 \ \mu$m.}
\label{diffgap}
\end{figure*}

In Fig.~\ref{diffgap}, we compare the density of excited electrons $n_{\rm ex}$ calculated with different bandgap values: the left and right panels show results for a pump-laser wavelength of 0.8 and 3.5 $\mu$m, respectively.
We record the corresponding slopes $m$ and values $\rm n_{ex}$ in Table \ref{t3}. Here, the linear fit is performed using 14 points for Keldysh, 3 for TDDFT, 10 or 8 for SBE calculations with a 0.8 or a 3.5 $\mu$m laser, respectively.

\begin{center}
\begin{table}[ht!]
\begin{tabular}{|l|c|c|c|c|}
\hline  
Method & $m_{0.8}$ & $n^{0.8}_{\rm ex} [m^{-3}]$ & $m_{3.5}$  & $n^{3.5}_{\rm ex} [m^{-3}]$ \\
\hline 
Exp.& - & $2\times 10^{25}$   & - & $2\times 10^{25}$  \\
Keldysh(2.8 eV) & 2.00  &$1.5859 \times 10^{25}$ & 8.27 & $6.1822\times 10^{21}$\\
Keldysh(3.3 eV) & 2.98  &$ 1.0161 \times 10^{23}$ & 8.27 & $ 7.0758 \times 10^{19}$\\
SBE(2.8 eV) & 1.47 & $1.3465\times 10^{25}$ & 1.97 & $6.3910\times 10^{25}$\\
SBE(3.3 eV) & 1.72 & 0.8171$ \times 10^{25}$ & 1.95 & $ 5.1618 \times 10^{25}$ \\
TDDFT(2.8 eV) & 1.89 &$2.0015\times 10^{25}$ & 1.26 & $2.1536 \times 10^{25}$  \\
\hline
\end{tabular}
\caption{Comparison of the slope $m$ and the density of excited electrons $n_{\rm ex}$ for different bandgap values, considering pulses wavelengths of 0.8 $\mu$m and 3.5 $\mu$m. }
\label{t3}
\end{table}
\end{center}

For a pump-laser with a wavelength of 0.8 $\mu$m (corresponding to an energy of 1.55 eV), a higher bandgap  hinders electron excitation, leading to a reduced value of $n_{\rm ex}$ for both the SBE and the Keldysh model.  In particular, two-photon absorption (2PA) can occurr if the band gap is 2.8 eV, while three-photon absorption (3PA) is always necessary for a bandgap of 3.3 eV. The transition from 2PA to 3PA mechanism results in an increase of the slope $m$. 

In left panel of Fig.~\ref{diffgap} we can observe for the Keldysh model that the purple circles (bandgap of 3.3 eV) always fall in the 3PA regime, while the red circles (bandgap of 2.8 eV) exhibit a transition from 2PA to 3PA at intensities of about $10^{17} \text{W/m}^2$. For higher intensities, both red and purple points are in the 3PA regime, with similar but not identical values of $n_{\rm ex}$. This discrepancy arises because the bandgap energy $E_g$ also influences the Keldysh parameter $\gamma$, which in turn affects $n_{\rm ex}$. In Fig.~\ref{diffgap}, the $\gamma$ values shown on the upper x-axis are calculated based on a value of the bandgap of 2.8 eV.
 
For a laser with a wavelength of 3.5 $\mu$m  (corresponding to 0.35 eV), $n_{\rm ex}$ also decreases when the bandgap is larger, but the slope $m$ for both SBE and Keldysh model remains nearly unchanged. This suggests that MPA is not the dominant mechanism in this case.
In the Keldysh model, the purple circles (bandgap of 3.3 eV) show a transition from 10PA to 11PA at an intensity close to $4 \times 10^{15} \text{W/m}^2$, causing a discontinuity indicated by a green rectangular frame. In contrast, the red circles (bandgap of 2.8 eV) remain within the 9PA regime in that range of intensities.
A further transition from 9PA to 10PA is visible for the red curve at about $10^{16} \text{W/m}^2$, while the purple curve shows a transition from 11PA to 12PA at a similar intensity.

The saturation of both curves at higher intensities is consistent with the decreasing value of the Keldysh parameter $\gamma$, that becomes smaller than 1, indicating the passage to the tunneling regime.

\section{Effect of longer propagation time}

As discussed in the main article, our simulated laser has a shorter propagation time than the real experimental pulse to reduce computational costs. To ensure reliability and enable a valid comparison with the experiment, we conducted an additional simulation with a longer propagation time using the "unmodified" propagation scheme (see the main text for details).

\begin{figure}[htbp!]
\centering
\includegraphics[width = 0.99\columnwidth]{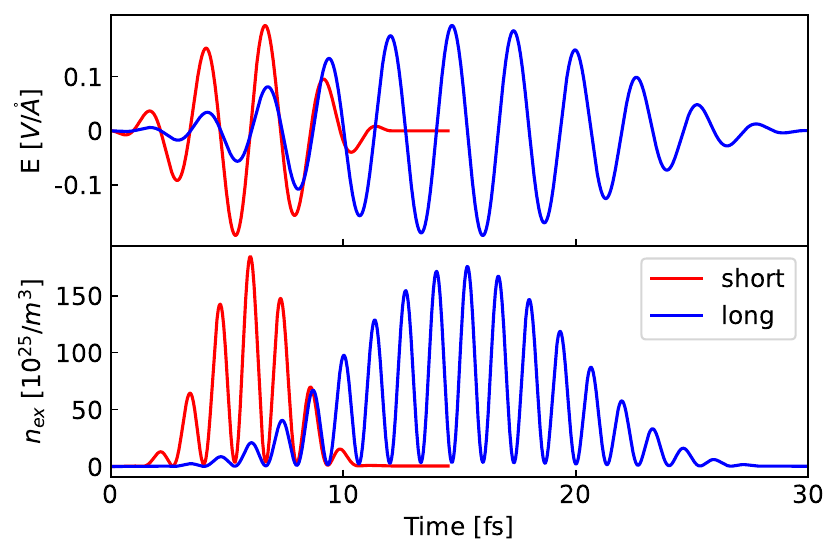}
\caption{(Upper panel) Electric field and (lower panel) excited electron density as a function of the propagation time.}
\label{shortlong}
\end{figure}

In Fig.~\ref{shortlong}, the pump laser has a wavelength of 0.8 $\mu$m and an intensity of $5\times 10^{15} W/m^2$ before entering the ZnO crystal.  The red curve represents the pulse used in our standard simulation, with a duration of 12.1 fs, while the blue curve corresponds to the experimental pulse with a propagation time of 30 fs.

After pulse propagation, the final density of excited electrons is very similar for both simulations: 
$n_{\rm ex} = 4.4\times 10^{24}/m^3$ at 12.1 fs for the short pulse and $n_{\rm ex} = 4.1\times 10^{24}/m^3$ at 28.5 fs for the long pulse. As the variation is minimal, we can validate the use of a shorter propagation time to reduce computational costs.

\bibliography{ref.bib}